\documentclass[prl,twocolumn,amsmath,amssymb,superscriptaddress,showpacs] {revtex4-2}
\usepackage{hyperref}
\usepackage{xcolor}
\usepackage{graphicx,psfrag,color}
\usepackage{dcolumn}
\usepackage{bm}
\usepackage{mathbbol, appendix}
\usepackage{printlen}
\usepackage{subcaption}
\usepackage[justification=Justified]{caption}
\usepackage{diagbox}
\usepackage{amsmath}
\usepackage{comment}
\usepackage[normalem]{ulem}
\usepackage{float}

\def\dual{\ \stackrel{\theta_\d}{\longrightarrow}\ }
\def\idual{\ \stackrel{\theta^{-1}_\d}{\longrightarrow}\ }
\def\qdual{\ \stackrel{\theta_\d (?)}{\longrightarrow}\ } 
\def\l{{\bm{l}}}
\def\d{{{\sf d}}}
\def\r{{\bm{r}}}
\def\x{{\bm{x}}}
\def\i{{\bm{e_1}}}
\def\j{{\bm{e_2}}}
\def\k{{\bm{e_3}}} 
\def\Z{{\mathbb{Z}}}
\def\one{{\mathbb{1}}}
\def\tr{{{\sf Tr}\ }}
\def\fg{\mathfrak{g}}
\def\pf{\mathcal{Z}}
\def\Z{{\mathbb{Z}}}

\newcolumntype{t}[1]{D{.}{.}{#1}}

\renewcommand{\selectlanguage}[1]{}
\begin{document}
\title{Universal fragility of spin-glass ground-states under single bond changes}

\author{Mutian Shen}
\affiliation{Department of Physics, Washington University, St.
Louis, MO 63160, USA}

\author{Gerardo Ortiz}
\affiliation{Department of Physics, Indiana University, Bloomington, IN 47405,USA}

\author{Yang-Yu Liu}
\affiliation{Channing Division of Network Medicine, Department of Medicine, Brigham and Women’s
Hospital and Harvard Medical School, Boston, MA, 02115, USA}
\affiliation{Center for Artificial Intelligence and Modeling, The Carl R. Woese Institute for Genomic Biology, University of Illinois at Urbana-Champaign, Champaign, IL 61801, USA}

\author{Martin Weigel} 
\affiliation{Institut für Physik, Technische Universität Chemnitz, 09107 Chemnitz, Germany}

\author{Zohar Nussinov}
\email{corresponding author: zohar@wustl.edu}
\affiliation{Rudolf Peierls Centre for Theoretical Physics, University of Oxford, Oxford OX1 3PU, United Kingdom}
\affiliation{Department of Physics, Washington University, St.
Louis, MO 63160, USA}
\affiliation{LPTMC, CNRS-UMR 7600, Sorbonne Universit\'e, 4 Place Jussieu, 75252 Paris cedex 05, France}

\date{\today}

\begin{abstract}
  We consider the effect of perturbing a single bond on ground-states of 
 nearest-neighbor Ising spin-glasses, with a Gaussian
  distribution of the coupling constants, 
  across various two and three-dimensional lattices and regular random graphs. 
  Our results reveal that the ground-states are strikingly fragile with respect to such changes. Altering the strength of only a single bond beyond a critical threshold
  value leads to a new ground-state that differs from the original one by a droplet 
  of flipped spins whose boundary and volume diverge with the system size --- an
  effect that is reminiscent of the more familiar phenomenon
  of disorder chaos. These elementary fractal-boundary \emph{zero-energy droplets} and their composites feature robust characteristics and provide the lowest-energy macroscopic spin-glass excitations. Remarkably, within numerical accuracy, the size of such droplets conforms to a universal power-law distribution with exponents dependent on the spatial dimension of the system. 
Furthermore, the critical coupling strengths adhere to a stretched exponential distribution that is predominantly determined by the local coordination number.
\end{abstract}


\maketitle

{\it Introduction.}
Complex systems harboring a plethora of competing low-energy states lie at the
forefront of intense investigation across diverse fields in physics, 
computation, biology, and network science (including longstanding foundational quests associated with the basic character of both real and
artificial neural networks and protein-folding)
\cite{SG1,Steinbook,Fortunatobook,Newmanbook,Bryngelson}. Spin-glasses are paradigmatic realizations of the venerable challenges posed by these
systems. Decades after their discovery, fundamental
aspects of spin-glasses \cite{Mydosh,SG1,Steinbook,Binder-Young} remain ill-understood. Excluding the fully-connected Sherrington-Kirkpatrick mean-field model \cite{SK} and
other soluble theories, e.g., ~\cite{REM,REM-long,Gross1984}, debates concerning the
nature of real finite-dimensional spin-glasses persist to this day. These systems are commonly described by the
nearest-neighbor Edwards-Anderson (EA) model \cite{EA}. 
We will take
the physically pertinent (and subtle) continuous real number
limit \cite{binomial} of the EA coupling constants prior to the thermodynamic
limit. \footnote{A delicate interplay exists between the latter two (continuum coupling and thermodynamic system size) limits. These two  limits do not commute with one another \cite{binomial}.}. 
With unit probability \footnote{Degeneracies arise for special values of the coupling constants (a set of measure zero).}, up to a trivial sign flip of all spins (a degeneracy henceforth implicit), the system provably has a unique ground-state \cite{binomial}. While some consensus emerged regarding the existence and character of the spin-glass phase-transition \cite{palassini:99,katzgraber:06a,hasenbusch:08}, at least in Ising systems, with
a lower critical dimension between 2 and 3 \cite{boettcher:06},
important questions remain regarding the spin-glass phase itself: e.g., 
whether there are asymptotically non-trivial 
overlap distributions and hierarchical structures of metastable states. A central engima is to what extent the alluring structure of the replica-symmetry breaking (RSB) solution of the mean-field
model survives in systems of finite dimensions $d$. Four
descriptions received 
much attention: (1) the full RSB framework
extended to finite dimensions \cite{parisi_replica_2012}, (2) the droplet scaling
theory \cite{mcmillan_scaling_1984,FisherHuse88,BM}, (3) the trivial-non-trivial
(TNT) 
\cite{TNT1,palassini_nature_2000,Houdayer2000}, and (4) the
chaotic-pairs (CP) pictures 
\cite{newman_metastate_1997,newman_ground_2021}. The most distinctive
features of these pictures relate to the relevant low-energy excitations. In the
RSB phase, such excitations have, asymptotically, an energy of order $O(1)$, independent of system
size, and space-filling domain-walls appear between pure-state regions. By contrast, 
conventional droplet scaling predicts energies $\sim \ell^\theta$ for excitations on scale $\ell$
with a fractal boundary of dimension $d_\mathrm{f} < d$
\cite{mcmillan_scaling_1984,FisherHuse88,BM}. The TNT and
CP scenarios feature $O(1)$, $d_\mathrm{f} < d$ (TNT) and high-energy ($\theta > 0$),
$d_\mathrm{f} = d$ excitations (CP)~\cite{newman_ground_2021}, respectively. In numerical studies,
such excitations are injected via boundary condition changes applied to systems
of linear size $L$ \cite{Cieplak83}. The corresponding ground-state energy scales as $L^{\theta}$ with $\theta$ 
negative in $d=2$ and positive when $d \ge 3$
\cite{BrayMoore84,HartmannYoung,Boettcher04}. However, since this setup requires a
non-local change of couplings, the resulting excitations might
not be representative of low-temperature behaviors. Several studies investigated local excitations \cite{TNT1,Houdayer2000,hartmann_generating_2004,Ard,PhysRevLett.86.3887} 
but their 
behaviors 
for short-range continuous spin-glasses 
remained somewhat inconclusive. 

\begin{figure}[tb]
  \centering
  \includegraphics[width = 0.45\textwidth]{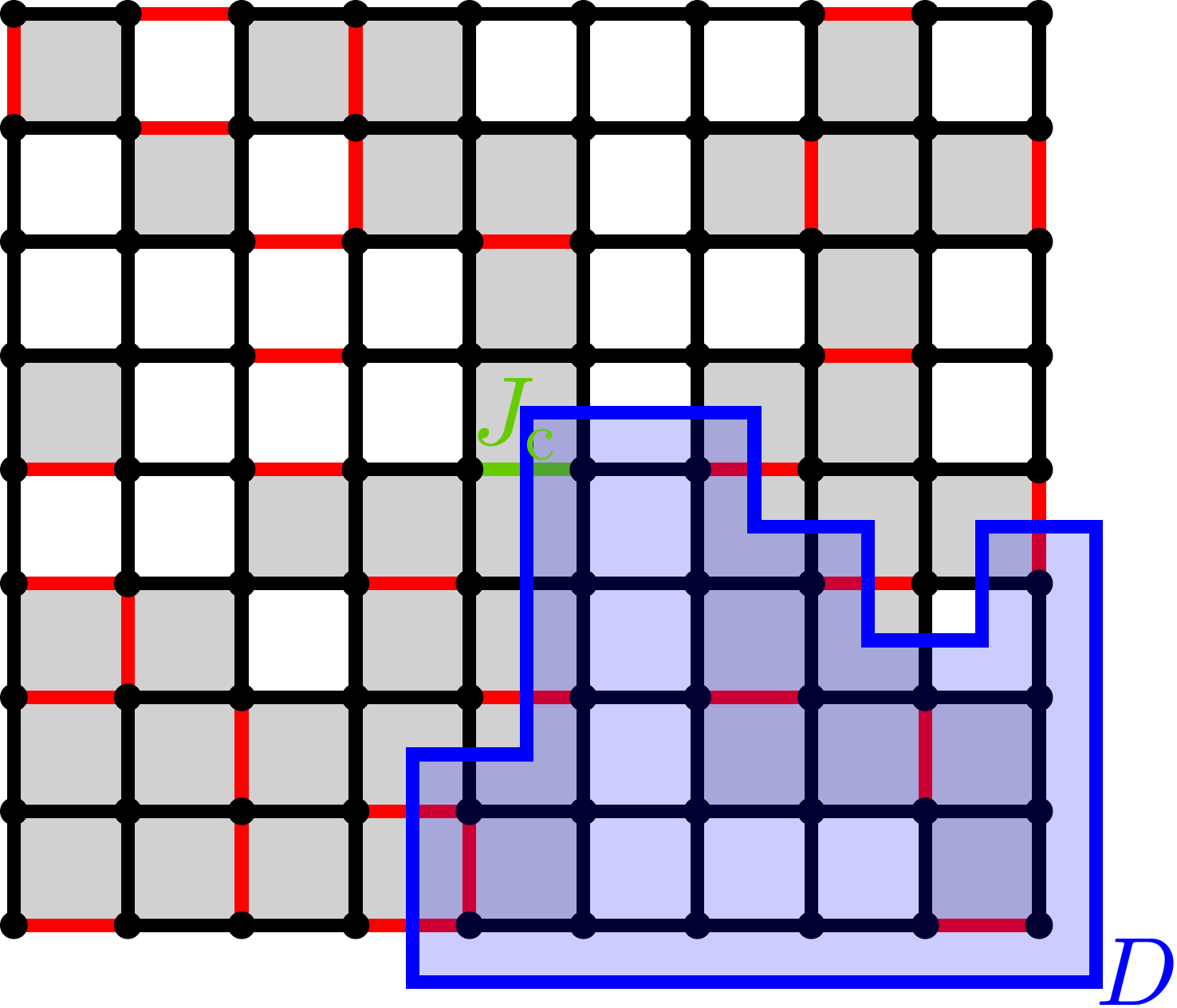}
  \caption{(Color online.)  Illustration of our numerical experiments. Gray and white squares represent frustrated and unfrustrated plaquettes, respectively. Red and black segments represent unsatisfied ($J_{ij}s_is_j<0$) and satisfied ($J_{ij}s_is_j>0$) bonds.
The critical value $J_\mathrm{c}$ of the coupling of the central bond $(i_0,j_0)$ (highlighted in green) separates
two different ground-states that differ by a domain $D$ 
    of flipped spins (shaded region). The boundary links in $\partial D$ are the bisected orthogonal (blue) edges. 
    }
  \label{fig:Lattice}
\end{figure}

{\it Zero-energy droplets.}
We consider the Gaussian EA
Ising model \cite{EA} with $N$ spins $\sigma_i=\pm 1$ and Hamiltonian
\begin{equation}
\label{Hamiltonian:eq}
	\mathcal{H}_J = -\sum\limits_{\langle i j \rangle}J_{ij}\sigma_i\sigma_j,
\end{equation}
where $\langle ij \rangle$ denotes 
nearest neighbors. 
 The couplings $\mathcal{J} = \{J_{ij}\}$ are drawn from a Gaussian
${\mathcal{P}}_J(J_{ij})=(1/\sqrt{2\pi})\exp(-J_{ij}^2/2)$. Starting from a ground-state of a given sample, we 
vary a {\it single}
coupling constant $J_{i_0j_0}$ of a bond $(i_0,j_0)$ at, e.g., the system center, from its initial strength $J_0$ until it reaches a critical value $J_{\mathrm{c}}$ at
which a new ground-state appears 
(see
Fig.~\ref{fig:Lattice}). 
Some properties of such droplets involving single
bond changes in the hypercubic EA Ising model \cite{hartmann_generating_2004} were studied analytically in Ref.~\cite{newman_ground_2021} yet specific results for the physically relevant 
cases in $d=2$ and $d=3$ were not provided. 
On tuning $J_{i_0,j_0}$, ground-states
become degenerate at $J_{i_0j_0} = J_\mathrm{c}$, differing by a 
domain of
flipped spins whose 
boundary is a contour of zero-energy. Previous work referred to the so-formed zero-energy droplet (ZED) as a
\emph{critical droplet} \cite{newman:00a}.  
Generally, spins flipped in any domain $D$ (
not necessarily 
a ZED) relative to those in the ground-state are associated with a boundary $(\partial D)$ 
energy \footnote{Here, $\sigma_k$ denotes the current spin at site $k$ following the spin inversion. The prefactor of two is a consequence of the sign inversion of the initial $\sigma_i\sigma_j$ following the spin flip in $D$.},
\begin{equation}
\label{positive_contour:eq}
	\Delta E 
 = - 2\!\!\!\! \sum\limits_{\langle ij \rangle \in \partial D} \!\!\!J_{ij}\sigma_i\sigma_j \ge 0 ,  \partial D=\{\langle ij \rangle | i \notin D, j \in D\}.
\end{equation}
The following properties can be proven \cite{SM,Arguin2019} 
(i) If $\Delta E =0$ (
a ZED), the set of 
flipped spins will contain exactly one of the two endpoints of the central bond \cite{newman_ground_2021}. (ii) As
$J_{i_0j_0}$ is continuously varied from $-\infty$ (where the central bond connects two oppositely oriented Ising spins) to $\infty$ (when the two spins are
parallel), there will only be a {\it single} ground-state transition at the critical
coupling $J_{i_0j_0}= J_\mathrm{c}$. Thus, if perturbing $J_{i_0j_0}$ to a new value generates a new ground-state ${\cal{C}}'$ then this state must differ from the original ground-state ${\cal{C}}$ by the very same 
spins in the ZED appearing when $J_{i_0j_0}=J_\mathrm{c}$ \cite{newman_ground_2021}. Furthermore, (iii) the energy associated with a ground-state
change (even if the number of flipped spins diverges)
incurred by altering a local exchange constant 
is asymptotically independent of system size if the distribution of the associated
critical couplings at which a transition occurs is well defined in the thermodynamic
limit. 
If the distribution of $J_\mathrm{c}$
values does not scale with system size (as we indeed verify) then neither will the
energy changes.

\emph{Multi-droplet excitations.}  We may vary couplings $J_{ij}$ on general (non-central) bonds and examine their respective ZEDs 
to study 
multi-droplet excitations for arbitrary ${\mathcal{J}}$. From \eqref{positive_contour:eq}, for general couplings, $\Delta E$ vanishes at ``criticality'' for non-trivial domains when degeneracy appears and is linear in deviations of the coupling
constants from their critical values.  
In the thermodynamic limit, for
a continuous distribution 
$\mathcal{P}_{J}$, one can find any finite number of disjoint bonds that are arbitrarily close to their critical values. 
  Thus, in that limit, the critical
boundary excitations that we examine and composites of a few such excitations may be of the lowest possible
energy. A related result holds for arbitrary energy excitations \cite{SM}. 

\emph{Numerical calculations}.  
We studied ZEDs by computing ground-states of the EA Ising model on (free boundary condition) square, triangular, and honeycomb lattices of linear size $16\le L\le 1024$, cubic lattices with $5\le L\le 12$, the body-centered cubic system with $L=5,\,7$, and
$z$-regular random graphs (RRGs) of coordination number $z=3$, $4$ and $6$ with $N=128$ nodes. 
  For each specific lattice or graph, we used between $10^3$ and $10^5$ bond configurations (disorder samples) for averaging \cite{SM}. For 
planar spin-glasses,
we used the polynomial-time minimum-weight perfect matching method \cite{bieche:80a,Kast} with \emph{Blossom V} \cite{kolmogorov_blossom_2009} to
determine exact ground-states.  
  For non-planar systems, an exact branch-and-cut approach
(implemented with \emph{Gurobi} \cite{gurobi_optimization_llc_gurobi_2022}) executes a brute force tree search of all possible spin configurations. 
The  code 
used for this work is publicly available \footnote{\url{https://github.com/renesmt/ZED}}.

\begin{figure}[tb]
  \centering
  \includegraphics[width=0.45\textwidth]{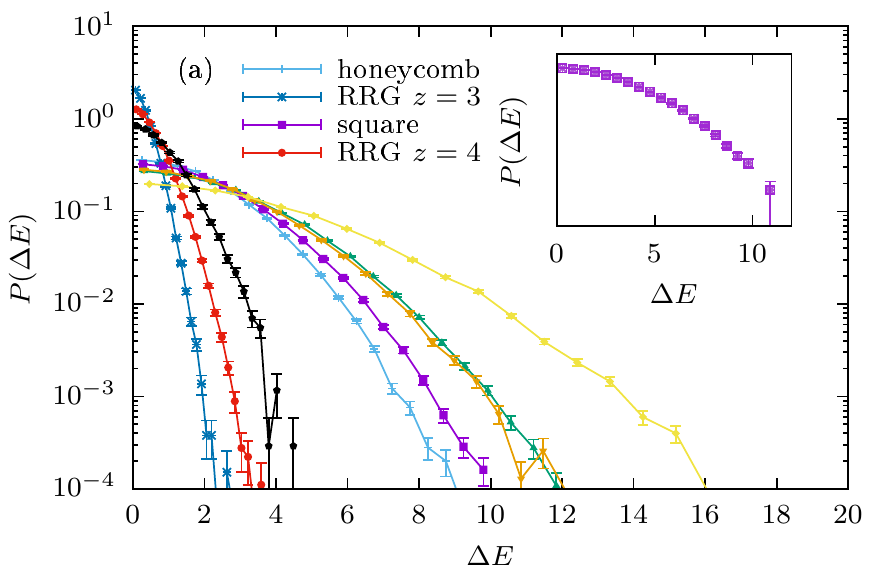}\\
  \includegraphics[width=0.45\textwidth]{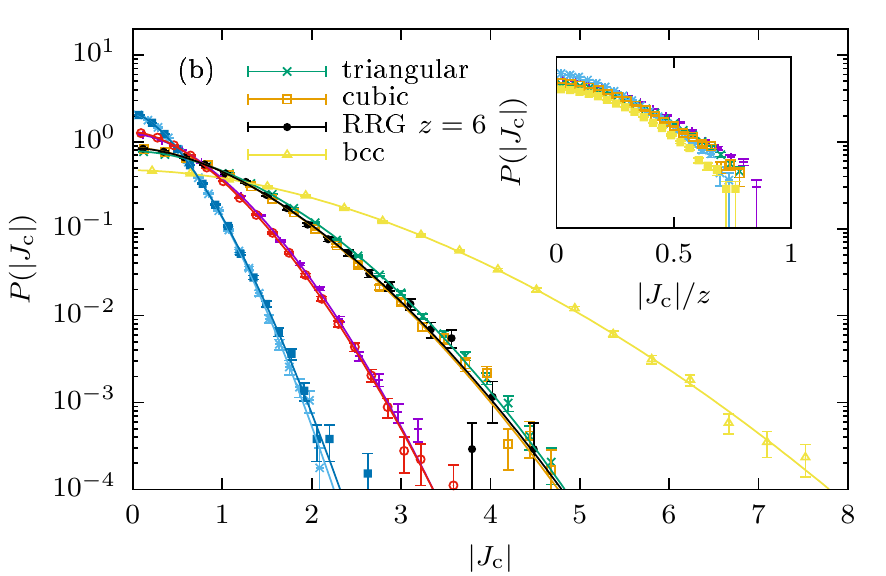}
  \caption{(Color online.) The legend is split between both panels, each of which shows all data sets. (a) Probability densities for the excitation
    energies $\Delta E$ 
    for 
    different lattices and graphs.  
    The
    inset shows the distributions for square lattices of sizes $L=16$, $64$, $256$ and
    $1024$ (darker shades for larger systems), illustrating that they are almost
    perfectly independent of system size.  (b) Probability densities of the modulus
    of the critical coupling, $|J_\mathrm{c}|$, together with fits of the stretched
    exponential form \eqref{pceq} to the data. Curves for lattices/graphs of the same coordination number $z$ are nearly indistinguishable. The inset shows the distributions as a
    function of $|J_\mathrm{c}|/z$.  }
  \label{fig:jcdistribution}
\end{figure}

\emph{Droplet energies and critical couplings.}
When the central coupling $J_{i_0j_0}$ varies from an initial strength of $J_0$ across $J_\mathrm{c}$ to a new value, the corresponding ground-state of \eqref{Hamiltonian:eq} transitions from ${\cal{C}}$ to 
${\cal{C}}'$. From the perspective of the original system with the initial coupling $J_{i_0j_0}=J_0$, the configuration ${\cal{C}'}$ constitutes an excitation of energy \footnote{This follows from Eq. (\ref{positive_contour:eq}) 
and property (ii).}
$\Delta E  = 2|J_0-J_\mathrm{c}|$ \cite{Arguin2019,SM} above that of the ground-state ${\cal{C}}$.
Using the latter relation for $\Delta E$, we inferred $J_\mathrm{c}$ by comparing the ground-states found for $J_{i_0j_0} \ll 0$ and $J_{i_0j_0} \gg 0$ respectively. 
 In Fig.~\ref{fig:jcdistribution}(a), we present excitation energies $\Delta E $ 
 for
the $d=2$ and $d=3$ lattices as well as the RRGs. The distributions are unimodal peaking
close to $\Delta E = 0$, with the energy changes increasing with the lattice (or graph) coordination number $z$. As the inset shows for the example of the square lattice, the distributions are almost perfectly independent of the lattice or graph size. Hence there is {\em no scaling of the excitation energies with
  system size\/}. To better understand these 
  distributions, we examined the behavior of the critical couplings $J_\mathrm{c}$. As their
probability density is even \cite{SM}, in
Fig.~\ref{fig:jcdistribution}(b) we show the distribution of the modulus
$|J_\mathrm{c}|$. These distributions are well described by a
stretched exponential (or stretched Gaussian) \cite{SM}

\begin{table}[tb!]
  \centering
  \caption{Parameters of the stretched exponential \eqref{pceq}, as well as   
  values of the scaling exponents $\kappa_\mathrm{v}$ of Eq.~\eqref{Fs} for the
    droplet volume and $\kappa_\mathrm{s}$ of \eqref{Ps} for the droplet boundary, for the different lattices
    considered. 
    }
  \begin{ruledtabular}
    \begin{tabular}{ct{6}t{6}t{6}t{6}}
      Lattice& \multicolumn{1}{c}{$a_\mathrm{c}$}& \multicolumn{1}{c}{$\beta_\mathrm{c}$} &
        \multicolumn{1}{c}{$\kappa_\mathrm{v}$} & \multicolumn{1}{c}{$\kappa_\mathrm{s}$}\\
      \hline
      honeycomb  & 2.76(1) & 1.58(1) & 0.215(2) &  0.342(3)    \\
      square  & 1.187(8) & 1.71(1)& 0.224(2) & 0.346(2)  \\
      triangular  & 0.523(7)  & 1.80(1) & 0.216(3)  &  0.336(3)  \\
      simple cubic  & 0.69(4)  & 1.55(5) & 0.131(6) &  0.159(5)\\
      bcc  & 0.21(1)  & 1.79(5)&  0.116(5) &  0.147(7)\\
    \end{tabular}
  \end{ruledtabular}
  \label{tab:k1d1}
\end{table}

\begin{equation}
  \label{pceq}
  P(|J_\mathrm{c}|) = k_\mathrm{c}\exp(-a_\mathrm{c}|J_\mathrm{c}|^{\beta_\mathrm{c}}),
\end{equation}
with $1<\beta_\mathrm{c}<2$. 
The lines in Fig.~\ref{fig:jcdistribution}(b) show fits of this
form with the parameters collected in Table \ref{tab:k1d1}. The typical values for
$J_\mathrm{c}$ are mostly determined by the lattice/graph coordination number $z$; the distributions almost collapse if plotted as a function of
$|J_\mathrm{c}|/z$, cf.\ the inset of Fig.~\ref{fig:jcdistribution}(b). For instance,
data for the ($z=6$) cubic lattice nearly collapse onto those of the ($z=6$)
triangular lattice. Similarly, the $P(|J_\mathrm{c}|)$ distributions for RRGs of fixed coordination $z=3$, $4$, $6$ but  
otherwise random
structure match with their counterparts of the honeycomb, square, and triangular
lattices respectively. 
Deviations are most pronounced for small $z$.
This is particularly apparent for a $z=2$ graph (a chain) for which $P(|J_\mathrm{c}|) =
\delta(|J_\mathrm{c}|)$ 
(since any sign change of 
$J_{i_0j_0}$ generates a new ground-state in which all spins on one side of this bond are flipped with degenerate 
ground-states at $J_{i_0j_0}=0$). Asymptotically, $P(|J_\mathrm{c}|)$ is independent of boundary conditions, although finite-size corrections might be strong \cite{SM}. 
We observed that the probability that the ground-state does not change when the initial central coupling is flipped ($J_{i_0j_0} \to -J_{i_0j_0}$) increases with the density of closed loops \cite{SM}.

\begin{figure*}[tb]
  \begin{center}
    \includegraphics[width=0.45\textwidth]{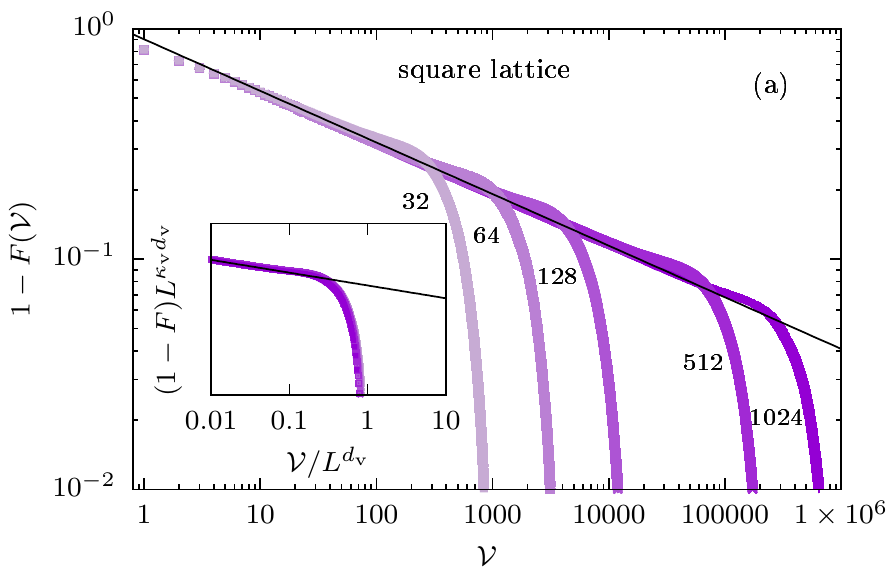}
    \includegraphics[width=0.45\textwidth]{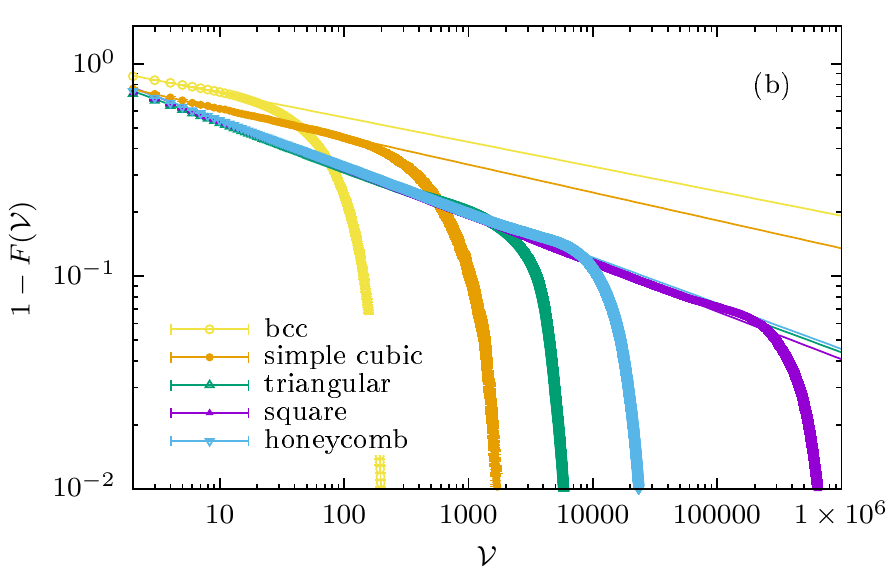} \\
    \includegraphics[width=0.45\textwidth]{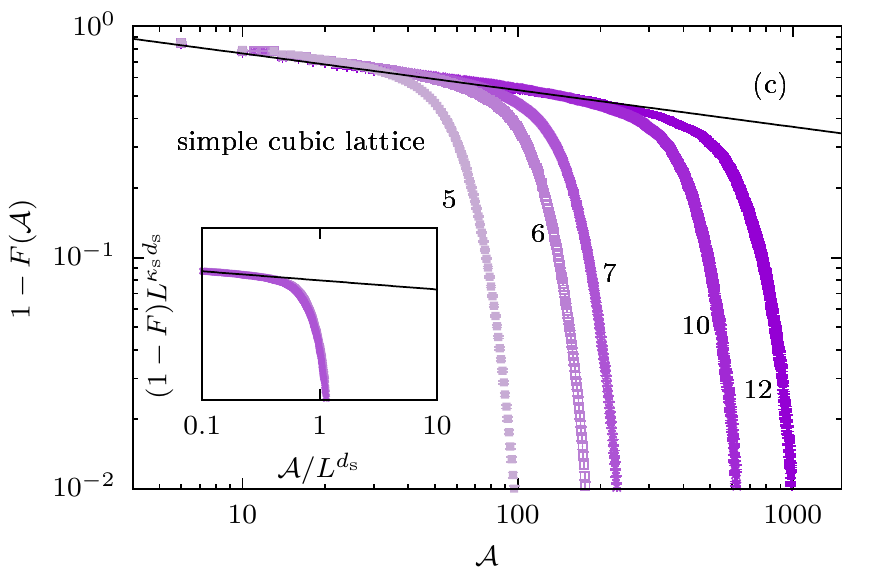}
    \includegraphics[width=0.45\textwidth]{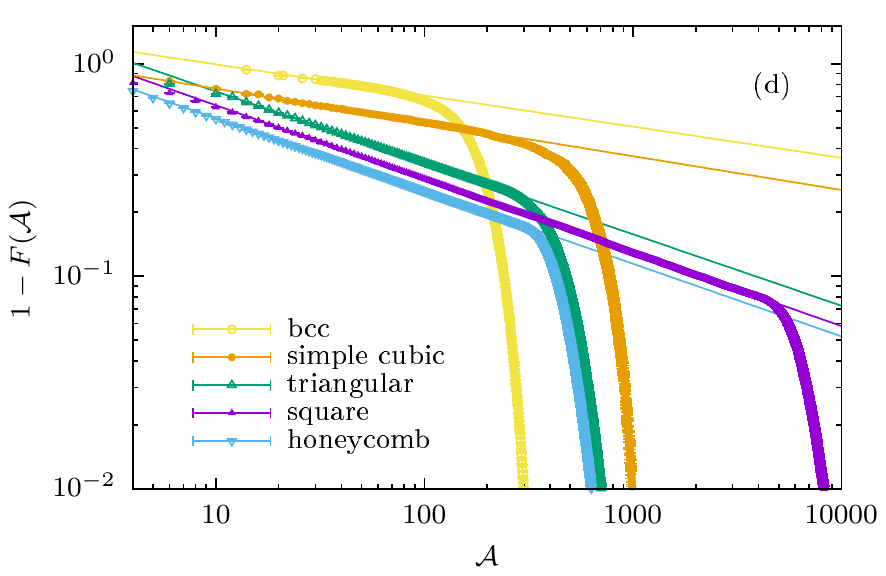}
  \end{center}
  \caption{(Color online.)
    ZED volume $\mathcal{V}$ and surface area $\mathcal{A}$ distributions. (a) Distribution 
    of ZED volumes for square lattices of sizes $L=32$--$1024$ (darker shades
    correspond to larger systems). The inset shows the scaling function $\Omega$ of Eq.~\eqref{Fs} assuming $\mathcal{V}_0 \sim L^{d_\mathrm{v}}$ with $d_\mathrm{v} = 1.991(75)$ (see below). (b) Volume distributions for the different lattice types with fits of
    the power-law form \eqref{Fs}. (c) 
    The ZED surface areas for
    simple-cubic lattices of sizes $L=5$--$12$. The inset shows $\Sigma$ of Eq.~\eqref{Ps} assuming $\mathcal{A}_0 \sim L^{d_\mathrm{s}}$ with $d_\mathrm{s} = 2.76(2)$ (see below). (d) Surface area distribution for different lattices with fits \eqref{Ps}. For the surface area distribution for the square lattices and the volume distribution for the simple cubic lattices, see Fig. S1 in the Supplemental Material.
  }
  \label{fig:CDF}
\end{figure*}

\emph{Droplet volumes and boundary areas.}
We next study the ZED
geometries. In Fig.~\ref{fig:CDF}(a), we show the tail distribution of the number of
sites $|D|$ (or volume) of these droplets  for square lattices of sizes
$32 \le L \le 1024$. All tails follow a power-law successively extending to larger droplet volumes, 
\begin{equation}
  P(|D| \ge \mathcal{V}) = 1-F(\mathcal{V}) = \frac{1}{\mathcal{V}_0^{\kappa_\mathrm{v}}} \Omega\left(\frac{\mathcal{V}}{\mathcal{V}_0}\right) \sim  k_\mathrm{v} \mathcal{V}^{-\kappa_\mathrm{v}},
  \label{Fs}
\end{equation}
where $F$ is the cumulative distribution and $\Omega$ a scaling function. Once the ZEDs become too
large, $\mathcal{V} \gtrsim \mathcal{V}_0(L)$, the finite size of the system becomes manifest
and the probability 
plummets far more rapidly with the ZED size. As is illustrated in Fig.~\ref{fig:CDF}(b), we find similar power-laws
for all considered $d=2$ and $d=3$ lattices. 
The 
exponent $\kappa_\mathrm{v}$
appears to only depend on the lattice dimension. 
Thus, we find the
compatible $\kappa_\mathrm{v} \approx 0.22$ for the square, triangular, and honeycomb
lattices, and $\kappa_\mathrm{v} \approx 0.125$ for the simple cubic, and bcc
lattices, respectively. The individual fit values appear in Table
\ref{tab:k1d1}. By comparison to their planar counterparts, 
the more notable differences between the simple cubic and bcc
lattices are likely a consequence of the smaller linear sizes in 
$d=3$. 

The ZED surface areas $|\partial D|$ exhibit a similar power-law distribution
\begin{equation}
  P(|\partial D| \ge \mathcal{A}) = 1-F(\mathcal{A}) = \frac{1}{\mathcal{A}_0^{\kappa_\mathrm{s}}} \Sigma\left(\frac{\mathcal{A}}{\mathcal{A}_0}\right) \sim k_\mathrm{s} \mathcal{A}^{-\kappa_\mathrm{s}}.
  \label{Ps}
\end{equation}
As seen 
in Fig.~\ref{fig:CDF}(c),
deviations from the power-law behavior occur for
$\mathcal{A} \gtrsim \mathcal{A}_0(L)$ with $\mathcal{A}_0(L)$ monotonically
increasing in $L$ \footnote{We 
  further monitor boundary effects and the cutoff $\mathcal{A}_\mathrm{0}$ when using open boundary
  conditions; on a square lattice, ZED perimeters of odd integer lengths are
  possible only for $\mathcal{A}> \mathcal{A}_\mathrm{0}$ (when ZEDs touch the boundary).}. As Fig.~\ref{fig:CDF}(d) illustrates, the exponents
$\kappa_\mathrm{s}$ are again universal, depending only on the lattice dimension,
cf.\ the fit parameters in Table \ref{tab:k1d1}. For RRGs with sparse closed loops,
  $P(|\partial D| \le \mathcal{A})$ becomes very sharp. For tree-like graphs (no closed loops),
  an entire branch of spins attached to the central bond flips when $J_{i_0j_0}$
  changes sign. Here, the boundary separating the ground-states for positive and
  negative $J_{i_0j_0}$ is comprised of only one ($|\partial D|=1$) bond and
  $P(|\partial D| \le \mathcal{A})$ increases sharply (a step
  function). Similarly, a higher exponent $\kappa_\mathrm{s}$ (sharper
  $P(|\partial D| \le \mathcal{A})$) appears for lattices of lower spatial dimension $d$ having
  fewer closed loops (see Table \ref{tab:k1d1}).


\emph{Fractal dimension.}  
Analyzing, for the square lattice, the scaling of the ZED volume
and surface areas with their linear extent $\ell$ \cite{SM}, we deduce a volume fractal dimension $d_\mathrm{v} = 1.97(3)$ and a surface fractal
dimension $d_\mathrm{s} = 1.27(1)$ \cite{SM}.
An alternative analysis using scaling collapses according to Eqs.~\eqref{Fs} and \eqref{Ps} yields the compatible estimates $d_\mathrm{v} = 1.991(75)$ and $d_\mathrm{s} = 1.275(30)$~\cite{SM}.
ZEDs are hence
compact (i.e., $d_\mathrm{v} \approx d=2$)  with fractal surfaces
of Hausdorff dimension compatible with that of 
domain-walls induced by changes
of the boundary conditions, $d_\mathrm{s,DW} = 1.2732(5)$ \cite{fractal2d}.
This similarity of fractal dimensions is intuitive as the flipping of boundary couplings
involved in transitioning from periodic to antiperiodic boundary conditions is akin 
to a sequence of ZED flips \footnote{As noted, property (ii) following Eq. (\ref{positive_contour:eq}), implies that (a) any non-empty domain $D$ of overturned ground-state spins in a new ground-state generated by changing a single bond, e.g., $J_{ij} \to -J_{ij}$, is identically the same as that of the ZED appearing when (b) $J_{ij}$ is tuned to criticality}. Indeed, the injection of a domain-wall can be viewed as sequentially flipping the bonds, one after the other, along a system boundary 
\cite{SM,newman_ground_2021}.
For the cubic lattice, a similar analysis yields $d_\mathrm{v}=3.08(5)[20]$ and $d_\mathrm{s}=2.76(2)[15]$, where the numbers in square brackets indicate the estimated systematic corrections from finite size~\cite{SM}. Since 
$d_\mathrm{v} \le d=3$, this suggests that 
$d_\mathrm{v} \approx d=3$. 
Again, 
$d_\mathrm{s}$ is comparable to previous estimates $d_\mathrm{s}\approx 2.6$~\cite{palassini_nature_2000}.

Given the cumulative power-law tail distributions of \eqref{Fs} and \eqref{Ps}, it is clear that the
probability \emph{densities} of volumes and surface areas decay algebraically (with
exponents $\kappa_\mathrm{v}+1$ and $\kappa_\mathrm{s}+1$, respectively) implying that the
{\em average\/} volume $\langle \mathcal{V} \rangle$ and surface area $\langle \mathcal{A} \rangle$ diverge with $L$.   
Specifically, the power-law ($\propto \mathcal{V}^{-(\kappa_\mathrm{v}+1)}$) regime 
of the ZED volume distribution implies
that 
$\langle \mathcal{V} \rangle > \int_{0}^{\mathcal{V}_0} d\mathcal{V}\,
\mathcal{V}^{-\kappa_\mathrm{v}}/\int_{0}^{\mathcal{V}_0} d\mathcal{V}\,
\mathcal{V}^{-(\kappa_\mathrm{v}+1)} \propto \mathcal{V}_0$. 
There are additional $\mathcal{V} > \mathcal{V}_0$ contributions not following the power-law 
\eqref{Fs}. 
The scaling collapse in the inset of Fig.~\ref{fig:CDF}(a) illustrates that
$\mathcal{V}_0 \sim L^{d_\mathrm{v}}$ such that the average volume diverges 
with $L$. Likewise, 
$\langle \mathcal{A} \rangle \sim L^{d_\mathrm{s}}$. Hence critical ZEDs are excitations of {\em divergent length scales\/} with {\em fractal boundaries\/}.

{\it Discussion.} The ground-states of the Gaussian EA Ising model are exceedingly fragile
and respond with (ZED) excitations of unbounded size to perturbations of single 
 couplings. We find universal exponents governing the geometrical size of these
excitations, the distribution of (``critical'') single couplings,
and energies. In the thermodynamic limit, many couplings are inevitably arbitrarily close to being critical so an
infinitesimal amount of energy may create macroscopic system-spanning
excitations. 
All 
excitations (domain-wall or other) may be associated with ZEDs that appear as exchange constants are sequentially tuned to values that they assume when these excitations arise \cite{SM}. 
The energies of system-spanning $d=2$ domain-wall excitations of length $\ell$ vanish as $\ell^\theta$ with $\theta = -0.2793(3) < 0$ \cite{fractal2d}. Thus, large $\ell$ domain-walls in $d=2$ asymptotically become ZEDs. 
When keeping the couplings fixed, spin configurations associated with (generally system-spanning) single bond ZEDs constitute excitations of energies that do not scale with $L$. In $d=3$ or whenever $\theta > 0$, domain-wall excitation energies diverge with increasing $L$ and are thus less relevant for 
low-temperature physics. In $d=2$ and $d=3$ lattices, 
ZED volume and surface area distributions follow {\it universal power-laws} with finite lattice cutoffs. 
ZEDs have compact volumes
with Hausdorff dimensions 
$ d_\mathrm{v} \approx d$ and 
fractal
boundaries $d-1 < d_\mathrm{s} < d$  consistent with 
domain-walls in
$d=2$. The ZED size monotonically  increases with external field \cite{SM}. 


Our setup for investigating ZEDs is complementary to that for
``disorder chaos'' wherein randomness is introduced globally by perturbing all couplings in the system  
\cite{BrayMoore87,middleton2001,krzakala_disorder_2005,katzgraber_temperature_2007,Dandan2012,Arguin2019,wang:15,chen_parisi_2017,baity-jesi_temperature_2021}. This
leads to an energy contribution proportional to $\ell^{d_\mathrm{s}/2}$. According to droplet theory, the relevant energy
scale is $\Delta E \propto \ell^\theta$, suggesting disorder chaos whenever $d_\mathrm{s}/2 > \theta$. By their nature, ZED 
perturbations (whose existence is guaranteed in the thermodynamic limit) are {\em always\/} relevant low-energy excitations. 

Since vanishing-energy and more general excitations are composites of ZEDs \cite{SM}, our findings carry important consequences. 
The defining ZED characteristics 
impose constraints on the properties of 
excitations in 
various 
pictures. Although finite-size corrections can be strong for spin-glasses, the
power-law exponents in Table \ref{tab:k1d1} clearly indicate divergent
droplet sizes in $d=2$ and $d=3$. The ZEDs 
are compact with fractal, but not space-filling, boundaries and O($1$) energy, 
thus differing from 
conventionally considered spin-glass excitations 
~\cite{SM},
and 
providing 
a test 
for 
comprehensive spin-glass theories. 

\begin{acknowledgments}
We thank Daniel Fisher, David A. Huse, Michael A. Moore, Daniel L. Stein, and Gilles Tarjus for discussions and correspondence. ZN is grateful to the Leverhulme-Peierls senior researcher Professorship at Oxford supported by a Leverhulme Trust International Professorship grant [number LIP-2020-014]. Part of this work was performed at the Aspen Center for Physics, which is supported by National Science Foundation grant PHY-2210452.
\end{acknowledgments}

\bibliography{ref.bib}
\clearpage

\def\theequation{S\arabic{equation}}
\def\thetable{S\arabic{table}}
\def\thefigure{S\arabic{figure}}

\def\dual{\ \stackrel{\theta_\d}{\longrightarrow}\ }
\def\idual{\ \stackrel{\theta^{-1}_\d}{\longrightarrow}\ }
\def\qdual{\ \stackrel{\theta_\d (?)}{\longrightarrow}\ } 
\def\l{{\bm{l}}}
\def\d{{{\sf d}}}
\def\r{{\bm{r}}}
\def\x{{\bm{x}}}
\def\i{{\bm{e_1}}}
\def\j{{\bm{e_2}}}
\def\k{{\bm{e_3}}} 
\def\Z{{\mathbb{Z}}}
\def\one{{\mathbb{1}}}
\def\tr{{{\sf Tr}\ }}
\def\fg{\mathfrak{g}}
\def\pf{\mathcal{Z}}
\def\Z{{\mathbb{Z}}}

\begin{center}
    \textbf{\Large Supplementary Materials}
\end{center}

In what follows, we expand on various aspects.

\section{Numerical Setting}
In Table \ref{tab:numerical_details} we provide the sizes and number of samples studied for each kind of lattice/graph investigated. In Table \ref{tab:vafitconf} we provide the lower and upper cutoff values for the fits to the tail distributions of $|D|$ and $|\partial D|$, respectively. We monitored the value of $\chi^2$ per degree of freedom (shown in in Table \ref{tab:chi2}) while adapting the range $[{\rm start},{\rm cutoff}]$ of each fit with the goal of minimizing the effect of finite-size corrections experienced by very small and very large clusters.

\begin{table*}[htb!]
\small
  \centering
  \caption{Linear system sizes $L$ for each lattice (resp.\ number of nodes for RRGs) and the corresponding number of samples.}
  \begin{ruledtabular}
    \begin{tabular}{ct{3}t{3}}
      Lattice& \multicolumn{1}{c}{$L$}& \multicolumn{1}{c}{$N_{\text{sample}}$} \\
      \hline
      honeycomb  & [16,32,64,96,128] & [10^5,10^5,10^5,85395,10^5]   \\
      square  & [16,32,64,96,128,512,1024] & [10^5,10^5,10^5,10^5,10^5,139974,99742]  \\
      triangular  & [16,32,64,96,128] & [10^5,10^5,10^5,10^5,101318]  \\
      simple cubic  & [5,6,7,10,12] & [118201,59865,49915,3987,4003] \\
      bcc  &  [5,7]& [59883,5960] \\
      RRG $z=3$ &  128& 92830 \\
      RRG $z=4$  &  128& 97994 \\
      RRG $z=6$  &  128& 14969 \\
    \end{tabular}
  \end{ruledtabular}
  \label{tab:numerical_details}
\end{table*}
\begin{table}[tb!]
  \centering
  \caption{Fit ranges for the tail distribution fits to the distributions of ZED volumes and areas.}
  \begin{ruledtabular}
    \begin{tabular}{ct{6}t{6}t{6}t{6}}
      Lattice& \multicolumn{1}{c}{$L$}& \multicolumn{1}{c}{volume start/cutoff}& \multicolumn{1}{c}{area start/cutoff}  \\
      \hline
      honeycomb   &128 & 20/128&20/128    \\
      square   & 1024& 20/512& 20/1024 \\
      triangular    & 128& 40/128 & 40/128   \\
      simple cubic    & 12 & 5/64&20/128\\
      bcc    & 7& 5/25 & 40/128\\
    \end{tabular}
  \end{ruledtabular}
  \label{tab:vafitconf}
\end{table}
All of our numerically provided {\it distributions} (such as those of the critical couplings and ZED volumes and areas) and their associated averages (e.g., average ZED volumes and areas and their fractal dimension scaling) were {\it computed over the sample realizations} of the spin-glass Hamiltonian of Eq. (\ref{Hamiltonian:eq}). As discussed in the main text, in these sample realizations, couplings $J_{ij}$ were drawn with probability density ${\mathcal{P}}_J(J_{ij})=(1/\sqrt{2\pi})\exp(-J_{ij}^2/2)$.
\begin{table}[tb!]
  \centering
  \caption{Mean-square weighted deviation per degree of freedom, $\chi^2$/d.o.f.\ for the fits to the distributions of ZED volumes and surface areas.
    }
  \begin{ruledtabular}
    \begin{tabular}{ct{6}t{6}t{6}t{6}}
      Lattice& \multicolumn{1}{c}{$|\partial D|$ }& \multicolumn{1}{c}{$|D|$} &
        \multicolumn{1}{c}{$J_{\rm C}$} \\
      \hline
      honeycomb $L=128$   & 1.14 & 0.90&1.27    \\
      square $L=1024$  & 1.10& 0.94& 0.96  \\
      triangular $L=128$   & 1.24& 1.00 & 0.90    \\
      simple cubic $L=12$   & 0.70 & 0.80&1.49\\
      bcc $L=7$   & 1.66& 0.37 & 1.83\\
    \end{tabular}
  \end{ruledtabular}
  \label{tab:chi2}
\end{table}

\section{Averages and Medians}
See Table \ref{tab:meanmedian} for the averages and medians of the volumes and areas of the ZEDs for different lattices.
\begin{table}[tb!]
  \centering
  \caption{Averages (left)/medians (right) of the volumes and areas of ZEDs for different lattices.}
  \begin{ruledtabular}
    \begin{tabular}{ct{6}t{6}t{6}t{6}}
      Lattice& \multicolumn{1}{c}{$|\partial D|$ }& \multicolumn{1}{c}{$|D|$} \\
      \hline
      honeycomb $L=128$   & 1886.4/15 &	105.7/14   \\
      square $L=1024$   & 26513.3/14& 692.3/20  \\
      triangular $L=128$   & 639.6/13	& 145.7/34    \\
      simple cubic $L=12$    & 240.9/48	&297.9/144\\
      bcc $L=7$   & 111.1/83	&314.8/333 \\
    \end{tabular}
  \end{ruledtabular}
  \label{tab:meanmedian}
\end{table}

\section{Collapse analysis for the volume and surface area distributions}

\begin{figure*}[tb]
  \begin{center}
    \includegraphics[width=0.45\textwidth]{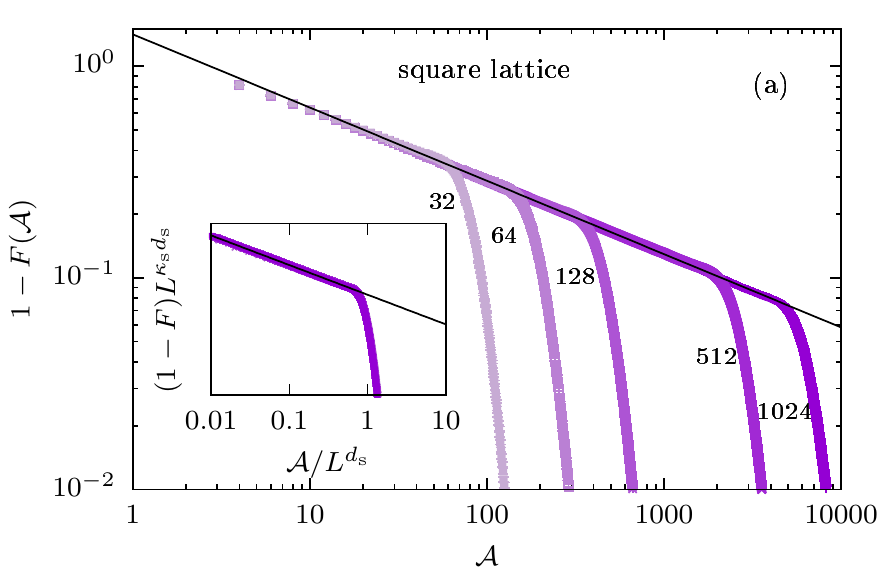}
    \includegraphics[width=0.45\textwidth]{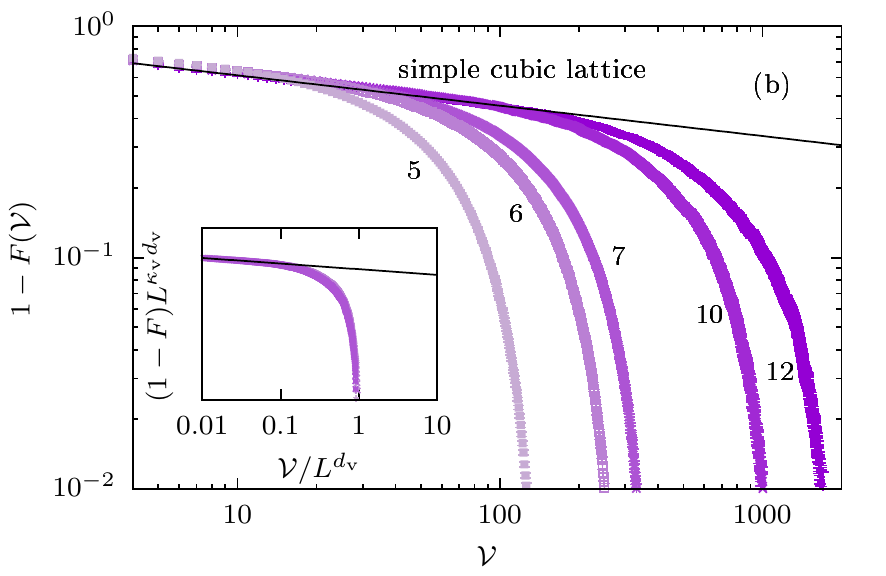}
  \end{center}
  \caption{(Color online.)
    (a) Tail distribution of ZED surface areas for square lattices of sizes $32 \le L \le 1024$. The inset shows the scaling function $\Sigma$ of Eq.~\eqref{Ps} assuming $\mathcal{A}_0 \sim L^{d_\mathrm{s}}$ with $d_\mathrm{s} = 1.275$. (c) 
    Tail distribution of ZED volumes for
    simple cubic lattices of sizes with $5 \le L \le 12$. The inset shows $\Omega$ of Eq.~\eqref{Fs} assuming $\mathcal{V}_0 \sim L^{d_\mathrm{v}}$ with the optimal collapse value of $d_\mathrm{v} = 3.08$. The quality of the collapse does not appreciably wane for a range of values due to systematic finite size error. Combined with the physical requirement that this volume  fractal dimension $d_\mathrm{v} \le d$, this  suggests that $d_\mathrm{v} \approx d =3$.
  }
  \label{fig:CDFsupp}
\end{figure*}

Figure 3 of the main text shows the distributions $P(|D|)$ and $P(|\partial D|)$ of the volumes $|D|$ and surface areas $|\partial D|$ of the ZEDs with a finite-size scaling analysis for $|D|$ and the square lattice in panel (a) and for $|\partial D|$ and the simple cubic lattice in (c). In Fig.~\ref{fig:CDFsupp} we complement these data by the distributions of $|\partial D|$ for the square lattice and $|D|$ for the cubic lattice.

\begin{figure*}[tb]
  \begin{center}
    \includegraphics[width=0.95\textwidth]{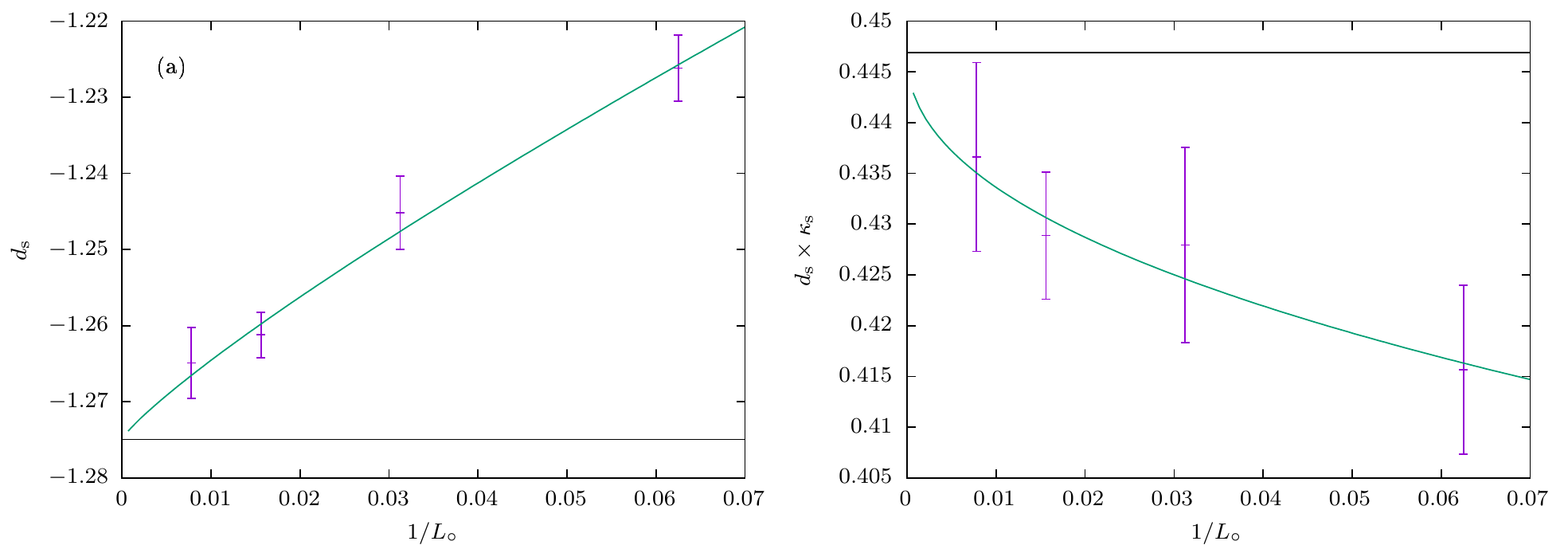}
    \includegraphics[width=0.95\textwidth]{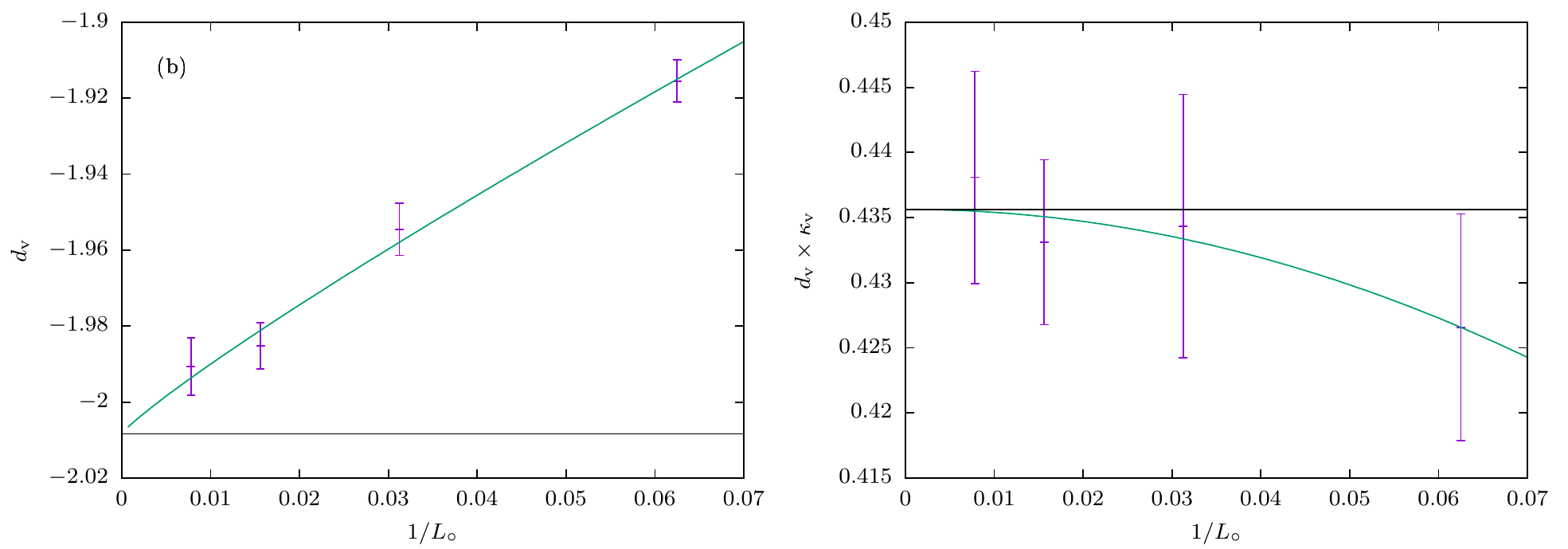}
\end{center}
  \caption{(Color online.)
    Extrapolation of the parameter estimates from the scaling collapses of the tail distribution of (a) the ZED volume and (b) the ZED surface area. The curved lines show power-law fits to the data, yielding the extrapolated estimates $d_\mathrm{v} = 2.008(58)$ 
    (and thus $d_\mathrm{v} \approx 2$ since, physically,  $d_\mathrm{v} \le d =2$) and $\kappa_\mathrm{v} = 0.2169(26)$ for the volumes and $d_\mathrm{s} = 1.275(29)$ and $\kappa_\mathrm{s} = 0.351(57)$ for the surface areas.
  }
  \label{fig:extrapolation-supp}
\end{figure*}

The insets of Figs.~3(a) and (c) as well as Fig.~\ref{fig:CDFsupp}(a) and (b) show collapses according to the scaling forms of Eqs.~\eqref{Fs} and \eqref{Ps} for the ZED volumes and surface areas, respectively. To determine the optimal collapse parameters $\kappa_\mathrm{v}$ and $d_\mathrm{v}$ resp.\ $\kappa_\mathrm{s}$ and $d_\mathrm{s}$, we made use of the collapsing tool \verb+autoscale.py+~\cite{melchert:09} that follows a procedure suggested in Ref.~\cite{houdayer:04}. To account for finite-size corrections, we performed the collapse for overlapping sequences of three lattice sizes, i.e., for $L=16$, $32$, $64$, then for $L=32$, $64$, $128$, followed by $L=64$, $128$, $512$, and finally $L=128$, $512$, $1024$ for the case of the square lattice. The resulting estimates are plotted against the smallest $L$ in the triple, $L_\circ$, in Fig.~\ref{fig:extrapolation-supp}. We then extrapolated the values of $\kappa_\mathrm{v}$ and $d_\mathrm{v}$ resp.\ $\kappa_\mathrm{s}$ and $d_\mathrm{s}$ in the limit $L_\circ\to\infty$, using a power-law form
\begin{equation}
  \alpha(L_\circ) = \alpha_\infty + a L_\circ^{-w},
  \label{eq:extrapolation}
\end{equation}
where $\alpha$ stands for one of the exponents $\kappa_\mathrm{v}$, $d_\mathrm{v}$, $\kappa_\mathrm{s}$, and $d_\mathrm{s}$, respectively. We find these fits to work well, resulting in values of $\alpha_\infty$ providing the estimates $d_\mathrm{s} = 1.275(29)$ and $\kappa_\mathrm{s} = 0.351(57)$, as well as $d_\mathrm{v} = 2.008(58)$ and $\kappa_\mathrm{v} = 0.2169(26)$. These are consistent with the values found from direct fits to the tail distribution as compiled in Table~\ref{tab:k1d1} of the main text. Physically, the volume fractal dimension $d_\mathrm{v} \le d$ thus suggesting that $d_\mathrm{v} \approx d=2$.

For the case of the simple cubic lattice, due to the much more limited system sizes $5 \le L \le 12$ the collapses are much less stable, and the resulting exponent estimates cannot be extrapolated using \eqref{eq:extrapolation}. We hence quote the collapse results for the largest triple $L=7$, $10$ and $12$ which are $d_\mathrm{v} = 3.08(5)[20]$, $\kappa_\mathrm{v} = 0.166(17)[8]$ and $d_\mathrm{s} = 2.76(2)[15]$, $\kappa_\mathrm{s} = 0.172(19)[4]$, respectively. Here, the numbers in square brackets are an estimate of the systematic (finite-size) error as implied by the variation of the collapse parameters for consecutive triples of system sizes. As the volume dimension $d_\mathrm{v}$ is bounded by the spatial dimension $d=3$, it follows that  $d_\mathrm{v} \approx d=3$. An estimate of the systematic error from comparing the collapse results for different ranges of system sizes leads to the final estimates of $d_\mathrm{v}$ and $d_\mathrm{s}$ provided in the main text. We note that the estimates for $\kappa_\mathrm{v}$ and $\kappa_\mathrm{s}$ from the collapses are consistent with those from direct fits as shown in Table~\ref{tab:k1d1}.

\section{ZED with Uniform Magnetic Field}
\begin{figure*}[tb]
  \begin{center}
    \includegraphics[width=0.45\textwidth]{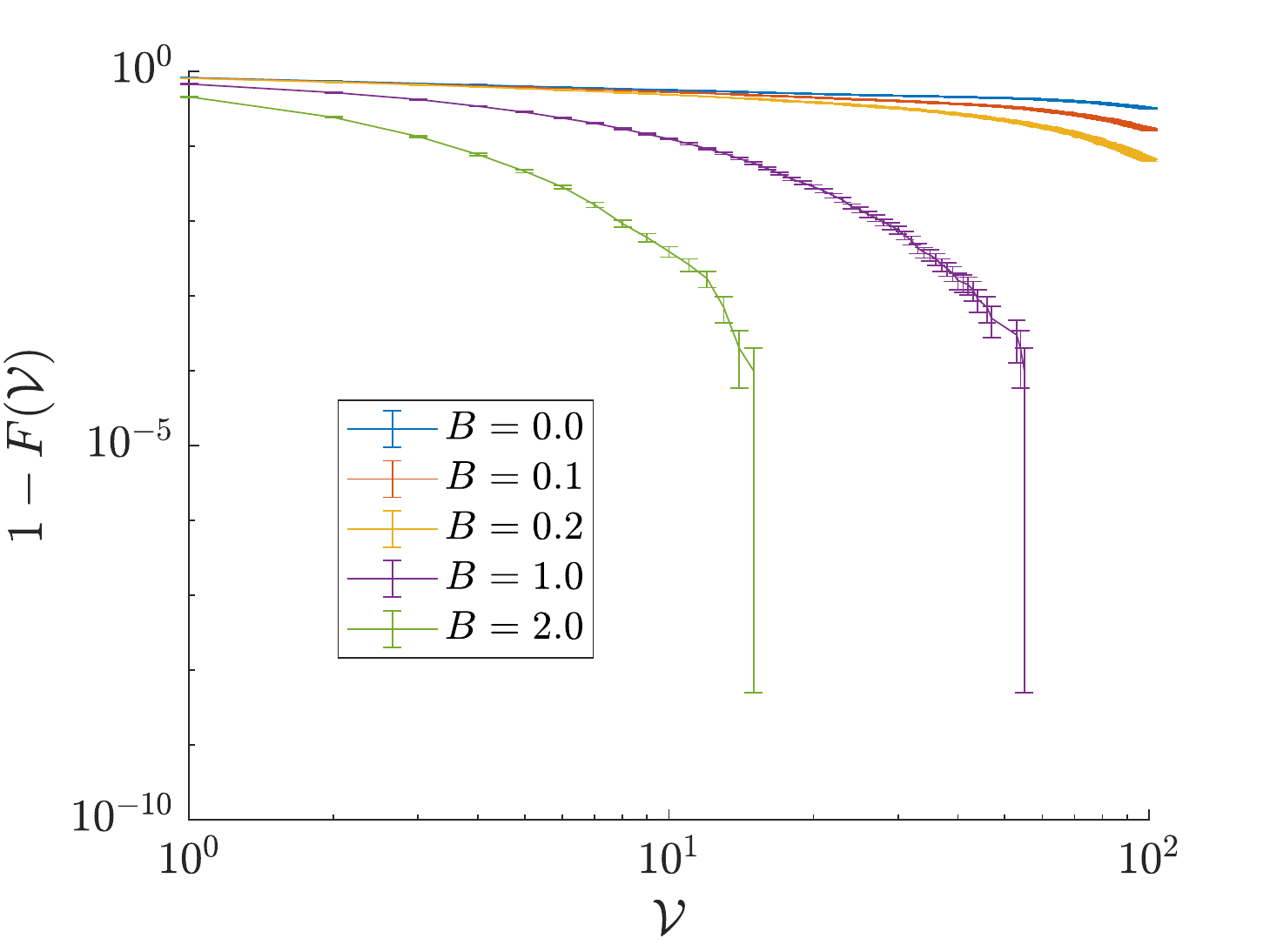}
    \includegraphics[width=0.45\textwidth]{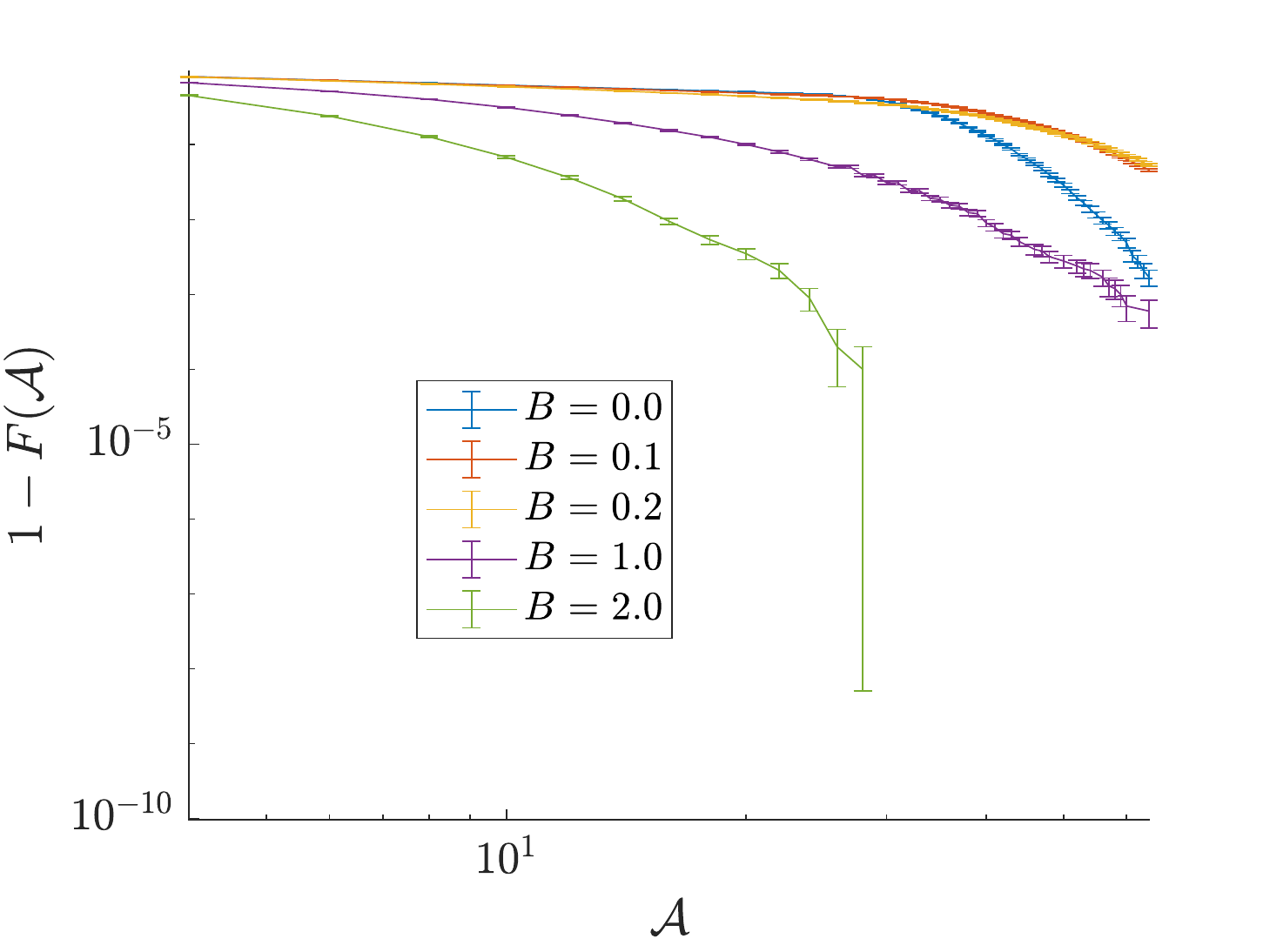}
  \end{center}
  \caption{The CDF distributions for the volumes (left) and areas (right) of ZEDs on $L=16$ square lattices for different applied uniform fields $B$. 
  }
  \label{fig:field}
  \end{figure*}

Applying a uniform magnetic field $B$ to the spin glass system (adding a term $B\sum_i \sigma_i$ to the Hamiltonian), the results are shown in Fig. \ref{fig:field}. In general, the ZED volume/area scales monotonically (decreasing in size) with increasing magnetic field strength $B$, see Fig. \ref{fig:field}.

\section{Topological Features of the ZED}

\begin{figure}
    \centering
    \includegraphics[width=0.48\textwidth]{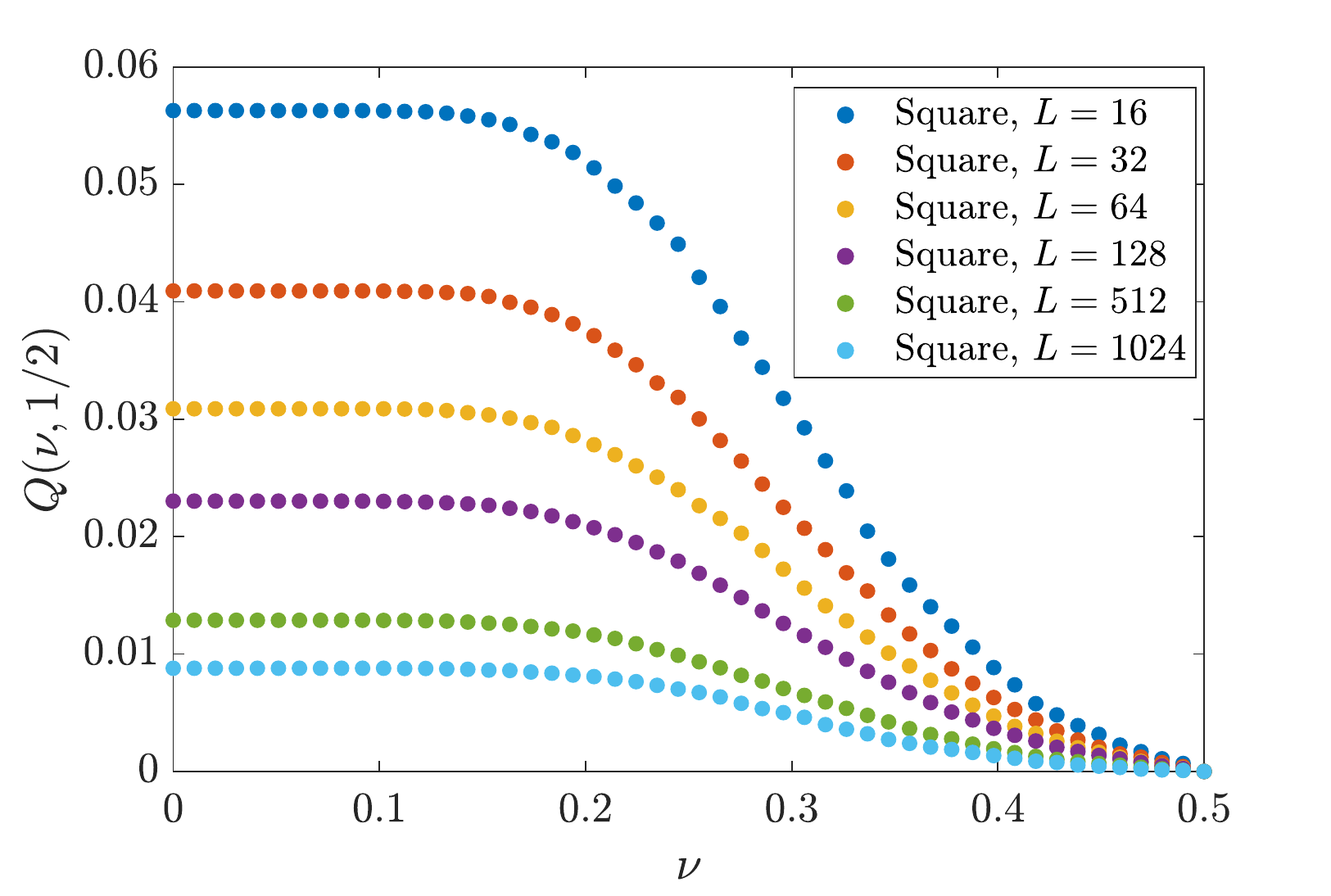}
    \includegraphics[width=0.48\textwidth]{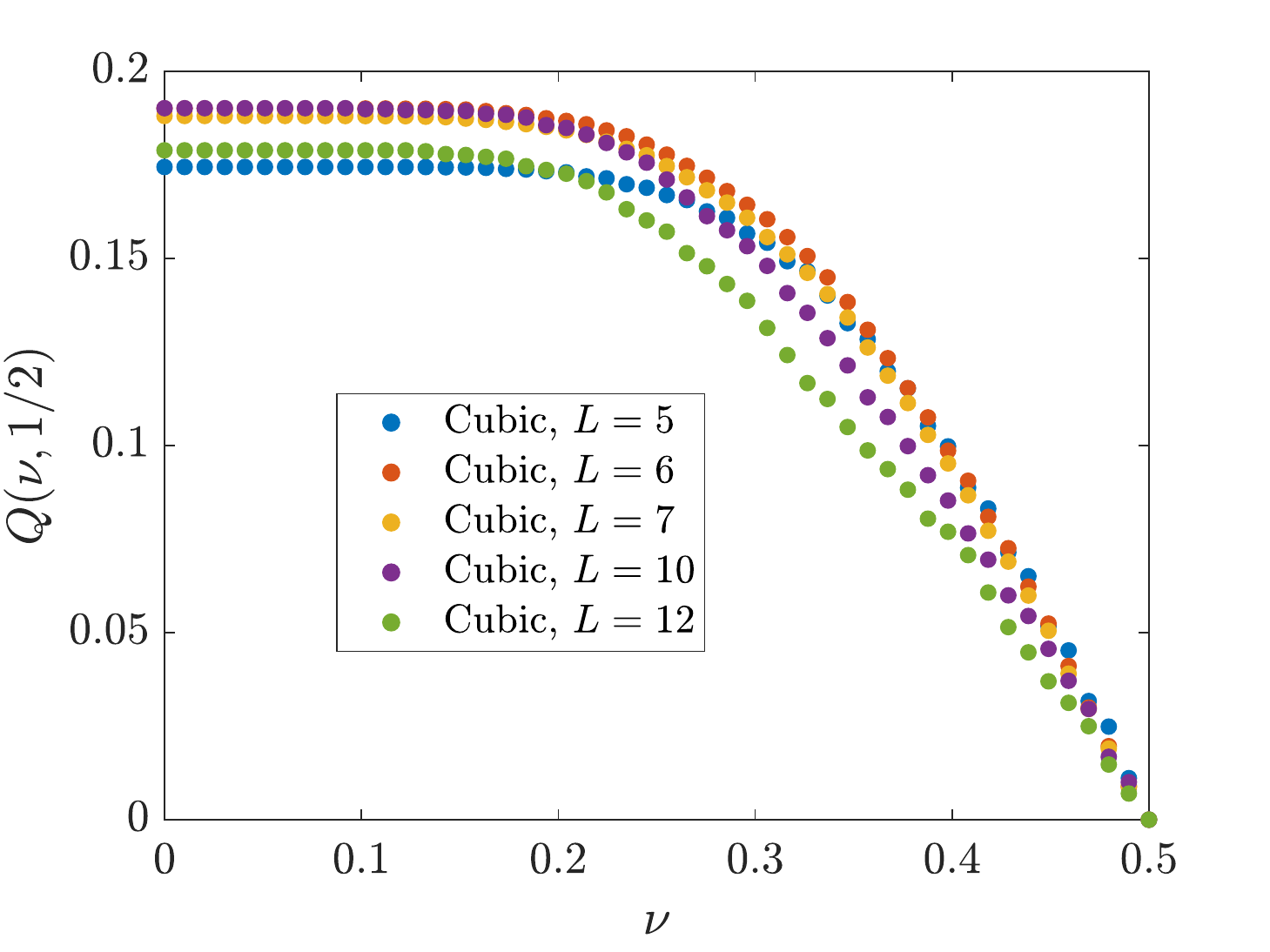}
    \caption{Here we show the probability $Q(\nu,1/2)$ of a ZED and its complement to touch all the boundaries of the system, for different values of $\nu$, i.e., so only ZEDs with volumes $|D|$ satisfying $\nu\leq |D|/N\leq 1/2$ are considered in each case.}
    \label{fig:topology}
\end{figure}

We studied some topological properties of ZEDs following the method proposed in Ref.~\cite{TNT1}. To this end, we considered the probability that the ZED and its complement (the lattice sites that are \emph{not} part of the ZED) both touch all boundaries in the system. For example, there are four line boundaries in the square lattices and six face boundaries in the cubic lattices. The volume $|D|$ of ZEDs is restricted to be at most half of the system volume $N$, i.e., $|D|/N\leq 1/2$, which is achieved by a global flip $\sigma_i\to -\sigma_i$ in case of a ZED occupying more than half of the sites. In order to show how these probabilities vary with the size of ZED, we plot the probabilities restricted to ZEDs of relative sizes $\nu\in [0,0.5]$, i.e., we only consider ZEDs with volumes $\nu \leq |D|/N\leq 1/2$. The results are shown in Fig. \ref{fig:topology} for the square and simple cubic lattices. There is a clear difference in the behavior between the 2D and 3D cases: while the probability of both the ZED and its complement touching the boundaries decays to zero as the system size $L$ is increased in 2D, it its largely independent of system size in 3D. In 2D there is simply not enough space for a cluster and its complement to percolate simultaneously. The behavior in 3D, on the other hand, illustrates the highly non-trivial topology of ZEDs that is reminiscent of the sponge-like structure proposed in the TNT picture ~\cite{TNT1}.

\section{Proofs of finite energy changes and other properties}
\label{PROOFS}
We now briefly prove properties (i)- 
(iii) of the main text. These properties are known and have been illustrated in other formats, e.g., \cite{Arguin2019,newman_ground_2021}. For the benefit of the reader, here we compactly demonstrate these important properties. Their proofs are quite simple. A violation of (i) implies that flipping all spins in $D$ (that does have the central bond as part of its boundary) creates a degenerate ground state at the critical coupling $J_{i_0j_0}=J_\mathrm{c}$. If $\partial D$ does not involve the central bond, this indicates that the original configuration with the unaltered $J_{i_0j_0}=J_{0}$ was degenerate with another ground state. However, apart from a set of vanishing measure in the coupling constants, the Gaussian EA Ising system is non-degenerate \cite{binomial}. Statement (ii) may be similarly proven by contradiction. Indeed, if additional transitions (apart from the one at $J_{i_0j_0} = J_\mathrm{c}$) between other degenerate ground states appeared in either of the two regimes in which both spins belonging to the central bond were either of the same or opposite relative orientation then these would only involve a change of other spins that do not belong to the central bond (thus violating (i)). To establish (iii), we recognize that although the ground state exhibits a change from a global spin configuration ${\cal{C}}^{-}$
at $J_{i_0j_0}< J_\mathrm{c}$ to a configuration ${\cal{C}}^{+}$ at $J_{i_0j_0} > J_\mathrm{c}$ (possibly involving a divergent number of spins having different assignments in the two respective ground states), the energy of any of the $2^{N}$ Ising spin configurations is, as a function of the coupling constants $\{J_{ij}\}$, linear. Thus, the minimum energy amongst these states is a continuous function of the coupling constants. Therefore, regardless of whether ${\cal{C}}^{-}$ and ${\cal{C}}^{+}$ differ by an extensive number of flipped spins (numerically, we find that the number of such flipped spins is indeed extensive), the ground-state energy is a trivial linear function of $J_{i_0j_0}$ as seen in \eqref{positive_contour:eq}. 

(A) If the initial and final values of the central coupling constant are both larger than $J_\mathrm{c}$ or if both are smaller than $J_\mathrm{c}$ then varying the central coupling constant between these initial and final values will not change the ground-state and thus the energy, as evaluated with the original value $J_{0}$ of the central coupling constant, will identically correspond to a vanishing energy change: $\Delta E=0$.  \newline

(B) We next briefly consider what transpires if the initial and final values of the central coupling constant lie on different sides of $J_\mathrm{c}$ (one being smaller than $J_\mathrm{c}$ and the other being larger than $J_\mathrm{c}$). The energy change can be readily computed with the aid of \eqref{positive_contour:eq} where now $D$ is the domain (ZED) of spins that differ in ${\cal{C}}^{-}$ and ${\cal{C}}^{+}$. By definition, precisely at criticality, the states ${\cal{C}}^{-}$ are ${\cal{C}}^{+}$ are degenerate and \eqref{positive_contour:eq} vanishes. As we vary the central coupling from $J_c$ to the value $J_0$ that it assumes in the initial Hamiltonian, only the term involving the central bond will change in the sum of \eqref{positive_contour:eq}. That is, 
$\Delta E$ will increase from its vanishing value at criticality as 
 $ \Delta E = 2|J_{0}-J_{c}|$. \newline

Thus, if distribution of $J_c$ values ($P(J_c)$) tends to well defined (system size independent) form in the thermodynamic limit (which numerically it indeed is (see the inset of Fig. \ref{fig:jcdistribution} (a)) then so is that of the associated energy changes ($P(\Delta E)$). As described in the main text, (B) was used to determine $J_\mathrm{c}$. While the statistical properties of $J_\mathrm{c}$ may depend on the system size for small systems, the latter energy change will always be of order unity (i.e., not increase with the system size since the coupling constants do not diverge with the lattice size). Similarly, as we further detail in this Supplementary Material, for the ``Repulsive Boundary Conditions'' there is a weak correlation between $J_\mathrm{c}$ and the associated ZED  boundary area  $|\partial D|$. We numerically find that the distribution $P(J_\mathrm{c})$ of critical coupling constants approaches a well-defined limit for large $L$. Given such a $P(J_\mathrm{c})$, both energy changes (A) (which does not depend on $J_\mathrm{c}$) and (B) will veer towards system size independent values for large $L$. That is, the distribution of the energy changes will tend to a unique form in the thermodynamic limit.

\section{Flipping the Sign of Central Bond and Probability of Changing the Ground State}
\label{sec:flip}
We now ask what will transpire if we flip the sign of the central bond $J_{i_0j_0}$ in the initial state (sampled from the Gaussian distribution of ${\cal{P}}_{J}$) rather than tune its value to the critical coupling value $J_\mathrm{c}$ at which the 
ground-state transition appears. 
Let us explicitly write the associated 
CDF as $F(J_\mathrm{c}) \equiv \int_0^{|J_\mathrm{c}|} P(|J'_\mathrm{c}|)dJ'_\mathrm{c}$. By construction, $F(J_\mathrm{c})$ is the probability that the critical coupling $J'_\mathrm{c}$ lies  within the interval $[-|J_\mathrm{c}|,|J_\mathrm{c}|]$. We may express the probability of changing the ground-state (G.S.) as
\begin{equation}
	\begin{aligned}
		P&({\text {G.S. Change}}) = \int^\infty_0 \mathcal{P}_J(J_{i_0j_0})~F(J_{i_0j_0})~dJ_{i_0j_0} \\
		&= \int^\infty_0 dJ_{i_0j_0} \frac{1}{\sqrt{2\pi}} e^{-\frac{J_{i_0j_0}^2}{2}}\int_0^{|J_{i_0j_0}|} dJ_\mathrm{c} ~~2P(|J_\mathrm{c}|).
	\end{aligned}
\label{eq:gschange}
\end{equation}
Eq. (\ref{eq:gschange}) relates the $J_\mathrm{c}$ distributions to the probabilities of changing the 
ground-states when we flip the central bond. In Table \ref{tab:prob}, we list the such probabilities for different types of lattices and different system sizes. Note the system sizes are different from what we have in Table \ref{tab:numerical_details}; this is another computation in which we  forcefully flip the bond, rather than tune it to $J_\mathrm{c}$.

\begin{table}[tbh]
	\centering
	\begin{tabular}{|c|c|c|c|}
		\hline
		& 8 & 28 & 48 \\
		\hline
  $d=2$ honeycomb & 0.2506 & 0.2525 & 0.2534 \\
  \hline
		$d=2$ square & 0.3795 &0.3744 & 0.3781
		 \\
		\hline
		$d=2$ triangular & 0.5110 &0.5133 & 0.5104 \\
		\hline
		& 3 & 5 & 7 \\
		\hline
		$d=3$ simple cubic & 0.4748 & 0.4887 & 0.4904 \\
		\hline
		& 2 & 3 & 4 \\
		\hline
		$d=3$ bcc & 0.6218 &  0.6604& 0.6685 \\
		\hline
	\end{tabular}
	\caption{The probability of changing the ground state $P({\text {G.S. Change}})$ following a flip of the sign of the initial random value of the central bond ($J_{i_0j_0} \to - J_{i_0j_0}
)$.}
	\label{tab:prob}
\end{table}
\section{The distribution $P(|J_\mathrm{c}|)$ of critical couplings for decoupled loops}
In what follows, we motivate a function that is  very similar (when $|J_c| \lesssim 1$) to that of Eq. (\ref{pceq})
for the particular (exactly solvable) case of decoupled loops of uniform fixed length. Towards this end, we define the conditional probability $Q(J) \equiv P(\sigma_i \sigma_j=+1|J_{ij}=J)$. Namely, $Q(J)$ is the probability that $\sigma_i \sigma_j$ assumes the value of $+1$ given that the link strength between the two spins is equal to $J$. Note that $Q(J)$ is, in fact, the cumulative distribution function of $J_\mathrm{c}$. That is, $Q(J)$ allows us to infer the probability that $J_\mathrm{c}<J$. That is, 
\begin{eqnarray}
\label{PQ}
P(J_\mathrm{c})=Q'(J_\mathrm{c}).
\end{eqnarray}
At the critical coupling strength $J_{\mathrm{c}}$,
the product $\sigma_i \sigma_j$ changes sign.

\begin{figure}
    \centering
    \begin{subfigure}[b]{0.49\textwidth}
    \centering
    \includegraphics[width=\textwidth]{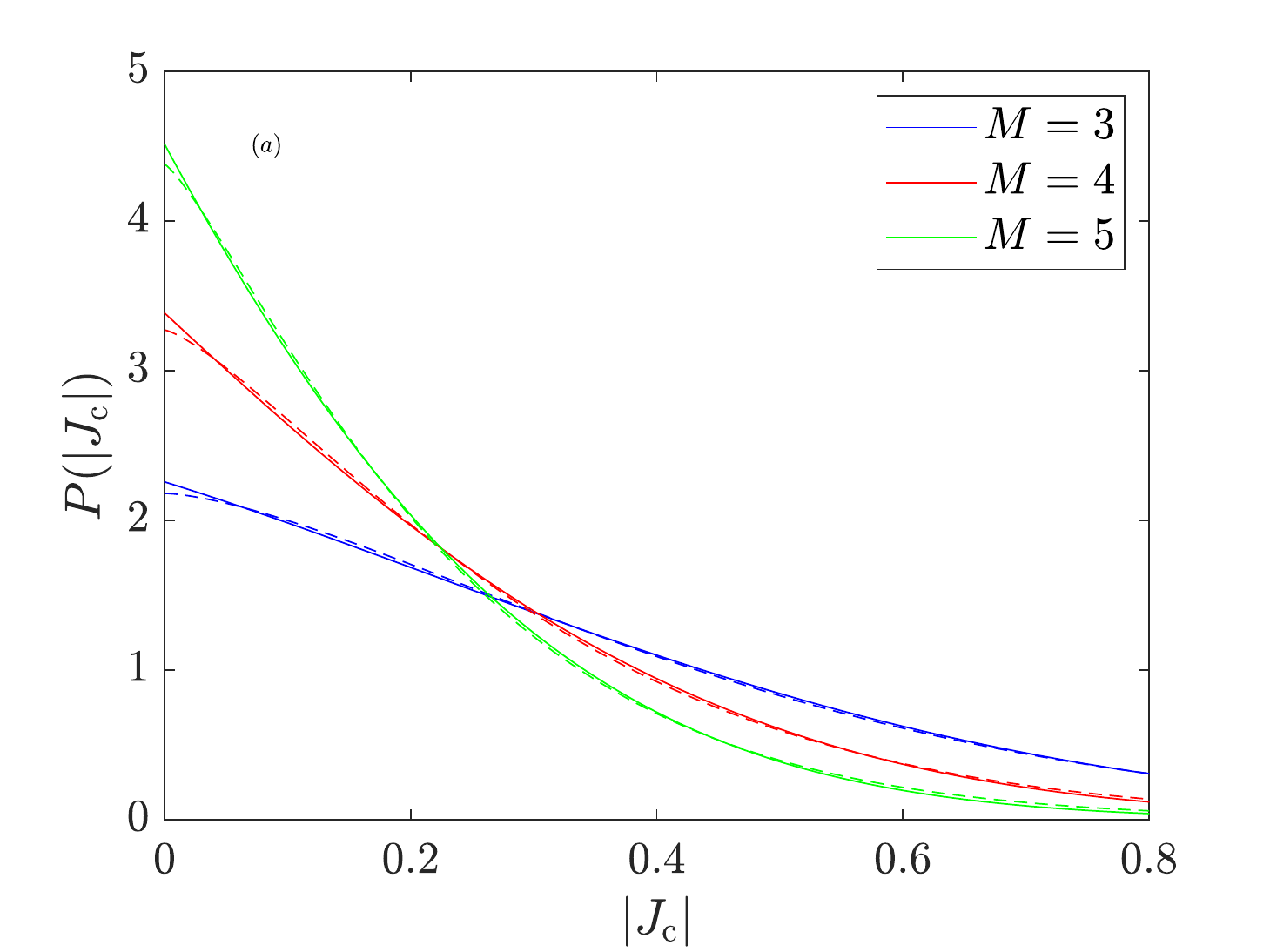}
    \end{subfigure}
    \begin{subfigure}[b]{0.49\textwidth}
    \centering
    \includegraphics[width=\textwidth]{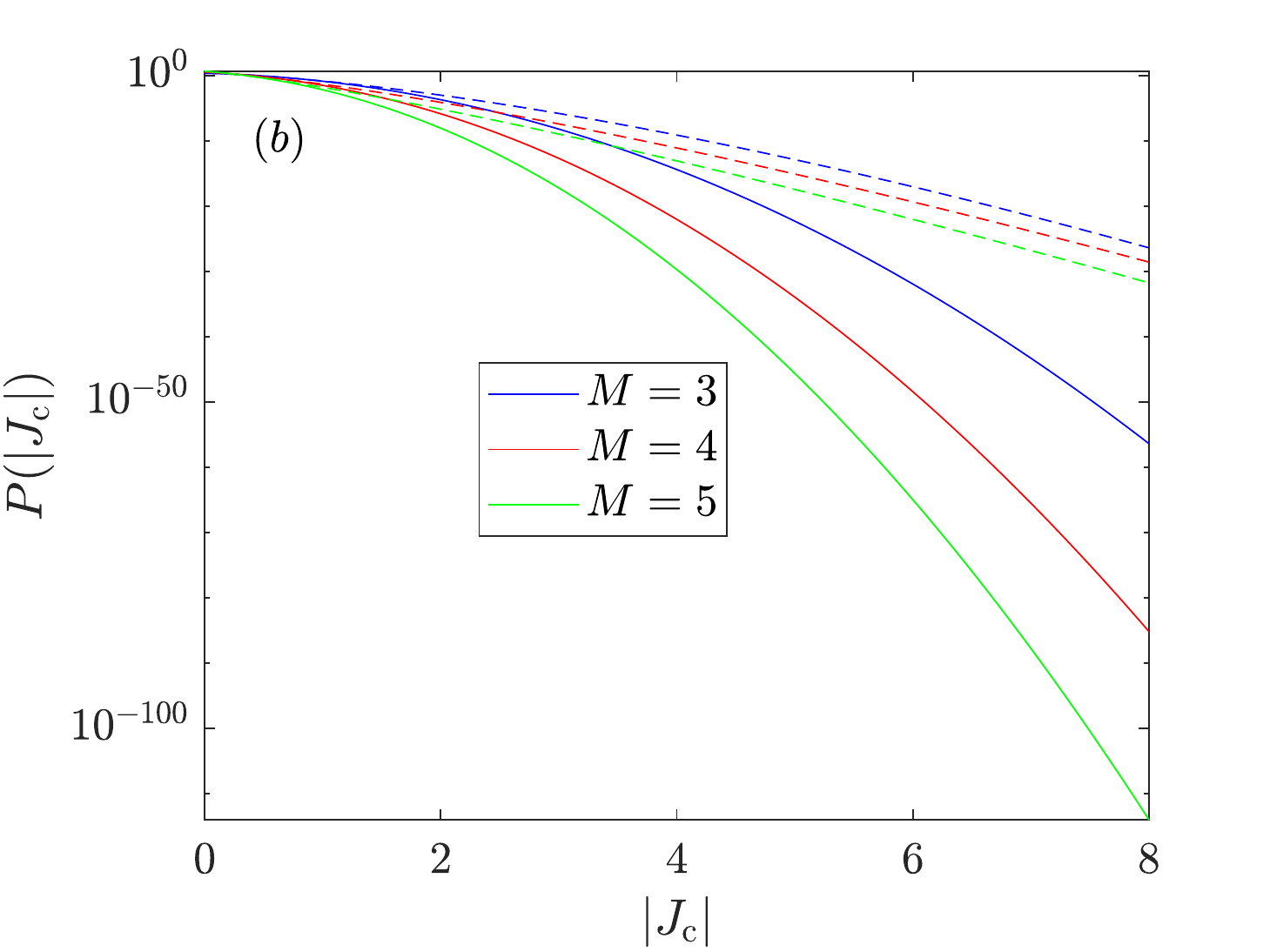}
    \end{subfigure}
    \caption{ (Color online.)
    The exact critical coupling distribution of Eqs. (\ref{PQ}, \ref{erfc}) (solid lines) for decoupled closed loops of length $M$ fitted by Eq. (\ref{pceq}) (dashed lines). (a) When $|J_c| < 0.8$
    and the probability density $P(|J_c|)$ is of order unity, one sees an exceptional agreement between the exact result of Eqs. (\ref{PQ},\ref{erfc}) for decoupled loops (dashed lines) and the more general numerical fit of Eq. (\ref{pceq}) (solid lines). This agreement is evident for all but the smallest values of $
|J_c|$. (b) When examined over the far broader range $|J_c| < 8$, there are very notable differences (on a logarithmic scale) between the exact result for decoupled loops and the numerical fit (of Eq. (\ref{pceq})) to that result. Note that the deviation shown is on a logarithmic scale. In terms of absolute size, both Eqs. (\ref{PQ}, \ref{erfc}) and well as the continuation of the empirical form of Eq. (\ref{pceq}) (when the latter is fitted in (a) to the exact result (Eqs. (\ref{PQ}, \ref{erfc})) for decoupled loops
for $|J_c| < 0.8$) to larger $|J_c|$ values are, literally, exponentially close to zero and should be completely unobservable. Empirically, however, the numerically measurable 
$P(J_c)$ is not nearly as small (see Fig. \ref{fig:jcdistribution}(b)). This discrepancy illustrates that the exact result for decoupled closed loops (Eqs. (\ref{PQ}, \ref{erfc}) misses essential features of the full lattice (and graph) systems that are correctly captured by Eq. (\ref{pceq}).}
    \label{fig:Q_derivative}
\end{figure}

We next investigate individual closed loops 
formed of $M$ individual bonds (e.g., $M=3$ for a single triangular plaquette, $M=4$ for a minimal square plaquette, etc.). Without loss of generality, for a particle bond $(ij)$, we consider $J_{ij}=J>0$. Whether 
closed loops are ``unfrustrated'' (i.e., all bonds in Eq. (\ref{Hamiltonian:eq}) can be simultaneously minimized) or ``frustated'' 
(when they cannot) depends on whether they respectively have an even or odd number of negatively valued couplings $J_{ab}$. For an unfrustrated closed loop, every single bond can be satisfied and thus for the specific link at hand, $\sigma_i \sigma_j=+1$. Within the ground-state of a frustrated closed loop, the link of the smallest absolute value $|J_{\min}| \equiv |J_{a^*b^*}| \equiv \min_{ab \in 
{\sf{loop}}}\{|J_{ab}|\}$ will be unsatisfied (that is for that link the product $J_{a^*b^*} \sigma_{a^*} \sigma_{b^*} <0$). Therefore, $\sigma_i \sigma_j=+1$ if and only if $|J_{ij}|$ is not the smallest bond among the $M$ bonds forming the closed loop, $|J| > |J_{\min}|$. Given that each of the $M$ links are drawn from the normal distribution ${\mathcal{P}}_J(J_{ab})=(1/\sqrt{2\pi})\exp(-J_{ab}^2/2)$, the probability that $J_{ij}$ is not the smallest coupling in absolute value amongst the $M$ coupling in the loop is, trivially, $1 - (1-  \frac{1}{\sqrt{2\pi}} \int_{-J}^{J} dJ' e^{-(J')^2/2} )^{M-1} $. Thus, for a single loop of length $M$,
\begin{eqnarray}
\label{erfc}
Q(J) = 1-(\mathrm{erfc}(J))^{M-1},
\end{eqnarray}
with $\mathrm{erfc}(J)$ denoting the complementary error function of $J$. Eq. (\ref{PQ}) then enables us to determine the probability distribution $P(J_c)$ of critical couplings. Numerically, when constrained to the interval $|J_c| \le 0.8$, except for exceedingly small arguments, the derivative of Eq. (\ref{erfc}) (i.e., $P(J_c)$ for decoupled loops) can be made match exceptionally well the empirical stretched exponential (or stretched Gaussian) distribution (Eq. (\ref{pceq}) of the main text), see Fig. \ref{fig:Q_derivative} (a). The difference between our exact result for decoupled loops with the fitted (for measurable probability densities) empirical general lattice and graph form of Eq. (\ref{pceq}) becomes far more acute when the $|J_c|$ interval is an order of magnitude larger. Indeed, as seen in panel (b) of Fig. \ref{fig:Q_derivative}, the critical probabilities predicted by Eq. (\ref{pceq}) (where the latter are fitted to  Eqs. (\ref{PQ}, \ref{erfc}) when $P(J_c)$ is finite and measurable) become exceedingly small for very large $|J_c|$. 
This disparity is not surprising. We must indeed underscore that the above derivation of Eq. (\ref{erfc}) applies for a single closed loop
and thus, by extension, only to decoupled loops of uniform length $M$. Clearly, the full problem on the lattice/graph in the thermodynamic limit is not that of the limiting case of decoupled loops of fixed length for which Eq. (\ref{erfc}) is exact. In other words, the scaling collapse of the $P(J_c)$ curves
in the inset of Fig.~\ref{fig:jcdistribution}(b)  highlighting the lattice/graph coordination number and the success of Eq. (\ref{pceq}) in describing RRGs for which only non-uniform sparse loops may appear does not, at all, follow from the above exact result for decoupled loops. It is for all of these reasons that we employed, in the main text, the more generic stretched exponential (or stretched Gaussian) fit of Eq. (\ref{pceq}) that numerically works well over the entire very broad numerically tested spectrum of $|J_c|$ values (see Fig.~\ref{fig:jcdistribution}(b)) and captures the coordination number dependence of $P(J_c)$ (see Fig. \ref{fig:jcdistribution}(b) and its inset).

\section{The Fractal Dimensions of Critical Droplets}
\label{A:fractal}
The fractal dimensions of ZEDs are defined from the scaling of the boundary area resp.\ the volume with its linear size.
The linear size of the droplet is defined as the `radius of gyration' in the same way as it is usually defined in the context of random percolation, see, e.g., Ref.~\cite{munster_cluster_2023}:
\begin{equation}
    \ell = \sqrt{\sum_{i=1}^s \frac{|\bold{r_i}-\bold{r_c}|^2}{s}},
    \label{eq:radius_of_gyration}
\end{equation}
where $s$ denotes the number of sites in the cluster, $\bold{r_i}$ is the coordinate vector of spin $i$ in the droplet, and $\bold{r_c}=\sum_{i=1}^s \bold{r_i}/s$ is the center of mass (note that we employ open boundary conditions).

We considered the droplets of a $1024\times 1024$ square lattice and droplets of a $12^3$ simple cubic lattice. The fractal dimensions were then extracted from the dependence of the average ZED volume and surface area on its linear extent. In computing the fractal dimension of the ZEDs, one should be aware of numerical details may influence the final result:
\begin{itemize}
	\item[1] The bin size $\ell_{\text{bin}}$. We have to bin the radii of gyration such that we can compute the corresponding average droplet size. We set the bin size to be $\ell_{\text{bin}}=4.0$ for 2D and $\ell_{\text{bin}}=0.1$ for 3D.
    	\item[2] The fit range $\ell_{\min} \leq \ell \leq \ell_{\max}$. We fixed $\ell_{\max}=140.0$ for 2D and $\ell_{\max}=1.5$ for 3D. We also tuned $\ell_{\min}$ within a certain range to see whether the fitted fractal dimension is stable to the fit range.
\end{itemize}
	
Employing such a numerical procedure, we
determine $d_{\mathrm{v}}=1.97(3)$, $d_{\mathrm{s}}=1.27(1)$ for 2D, while the values do not appear to settle down for the available system size range in the cubic lattice system.

\begin{figure}
    \centering
    \includegraphics[width=0.49\textwidth]{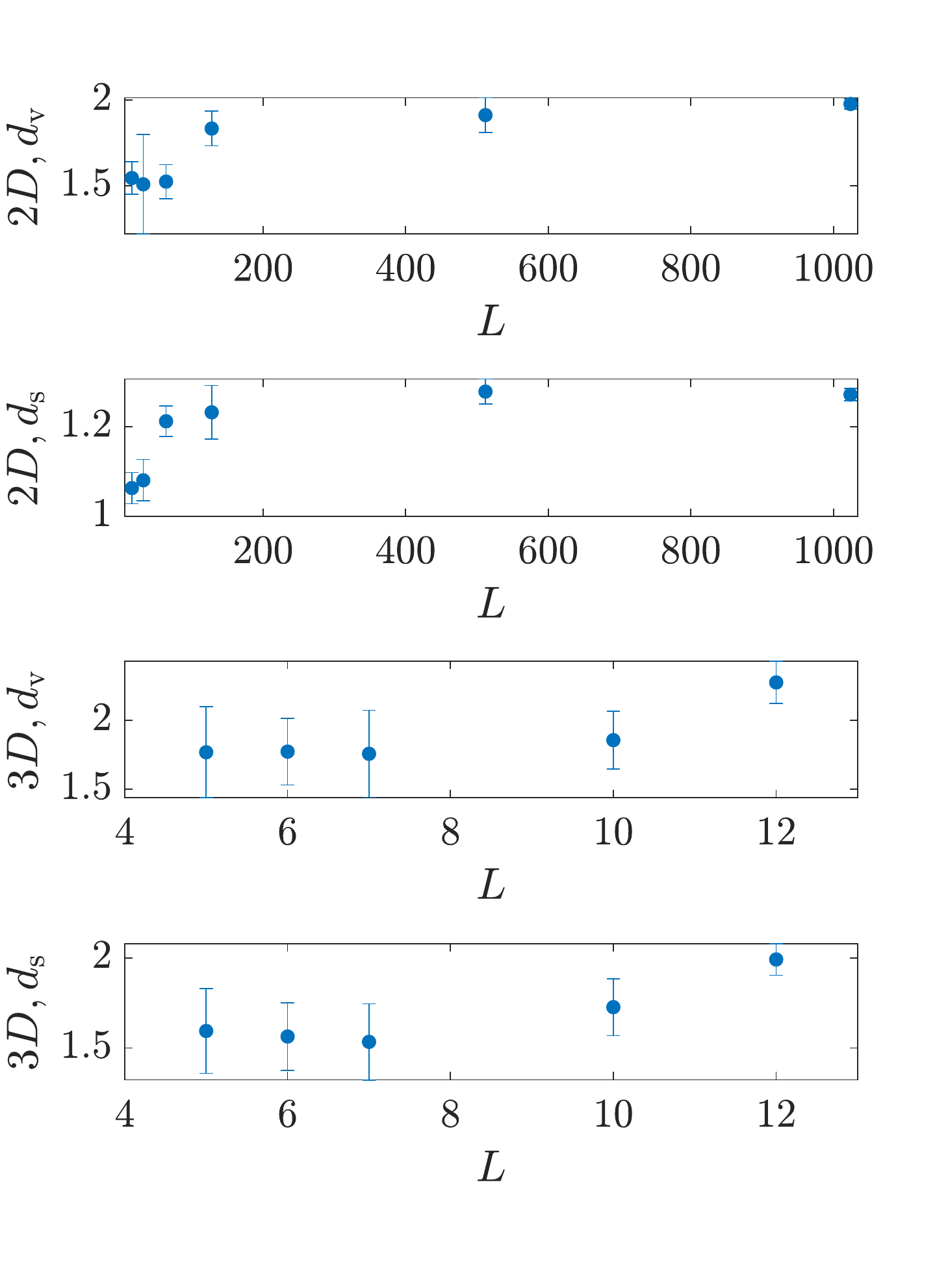}
    \caption{The fractal dimensions $d_{\mathrm{v}}$ and $d_{\mathrm{s}}$ estimated from the scaling of $|D|$ and $|\partial D|$ with the radius of gyration for different lattice sizes.}
    \label{fig:fracdim}
\end{figure}

The values in two dimensions are consistent with those extracted from the collapse analysis. The values in three dimensions, on the other hand, are likely to be strongly affected by finite-size corrections as the systematic upward trend in Fig.~\ref{fig:fracdim} shows. Unfortunately these results are not accurate enough for a systematic finite-size extrapolation. In contrast, the above discussed estimates from the collapse analysis are more stable, and we suggest that they are the more reliable estimates.


\section{Repulsive and half periodic boundary conditions}
\label{sec:Repel}
\begin{figure}[tbh]
	\centering
	\includegraphics[width=0.5\textwidth]{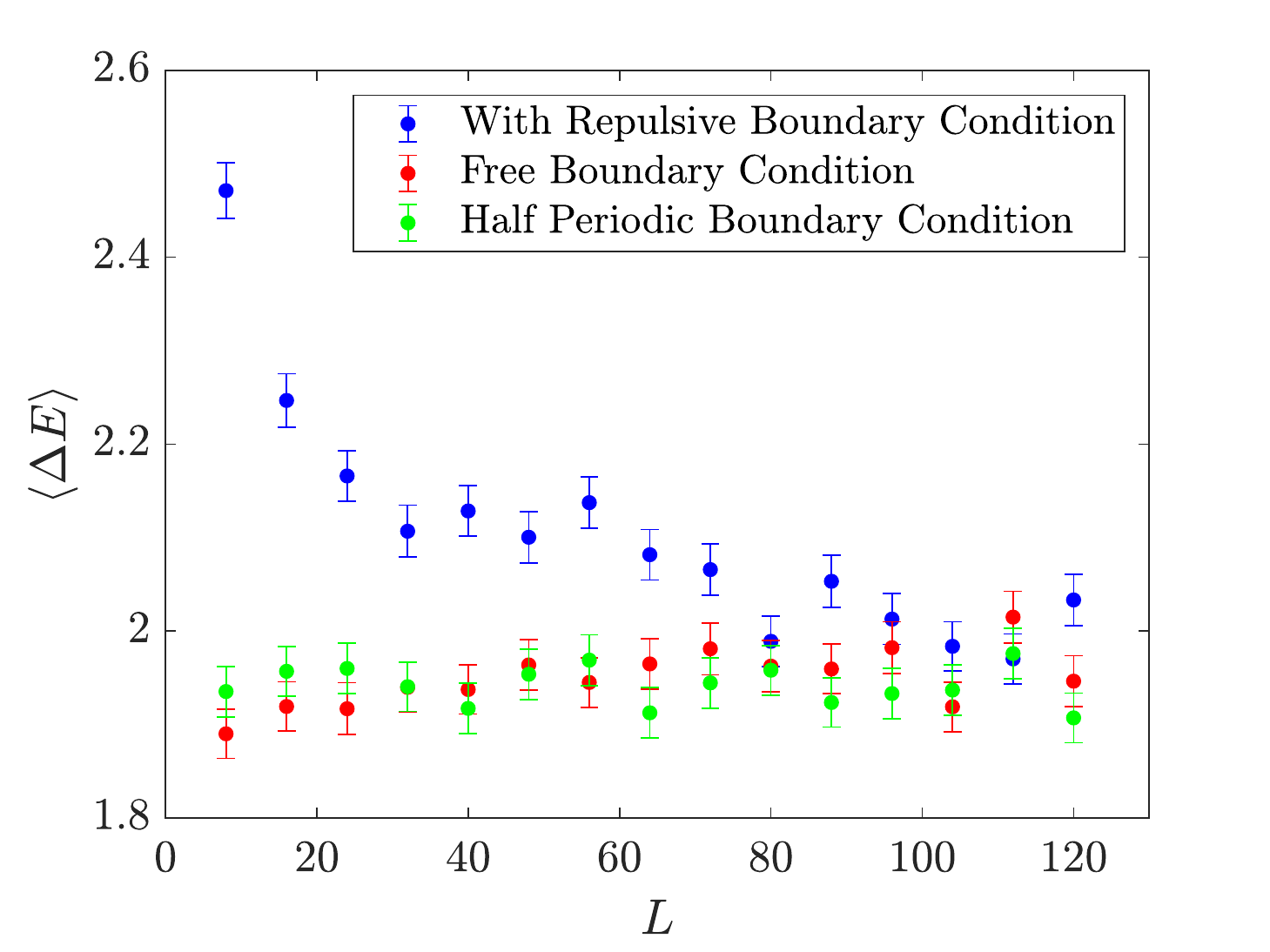}
	\caption{(Color online.) The average single bond average excitation energy $\langle \Delta E \rangle$ on $L \times L$ square lattices for (a) free boundary conditions, (b) half-periodic boundary condition and, (c) the repulsive boundary conditions (see text). For both the free boundary condition and half-periodic boundary condition, the average energy does not change systematically with the system size. For the repulsive boundary conditions, there is a nontrivial dependence of the energy on system size for small lattices before the energy difference appears to saturate (within numerical accuracy) to the very same value ($\langle \Delta E \rangle \simeq 2$) obtained with open and half-periodic boundary conditions. Note that this particular approximate limiting average value for the square lattice assumes other values for other lattices or graphs (see Fig. \ref{fig:jcdistribution} (a) for the full distributions).}     
	\label{fig:compare_rbc}
\end{figure}

We now discuss the role of boundary conditions in some detail. In the results that we reported on in the main text, we used open boundary conditions. For comparison with earlier work \cite{hartmann_generating_2004}, however,  we also employed 
what we call `repulsive boundary conditions’. Herein, the boundary spins are kept fixed so that, by construction, the droplets cannot reach (and are effectively repelled by) the boundary. Our analysis reveals that the divergent (in size) ZEDs and the exponents that we find might be more readily missed, due to finite size effects, by the use of these boundary conditions. Perusing ours and earlier numerical results with these boundary conditions \cite{hartmann_generating_2004}, one finds that if only the largest system sizes are employed with droplets lying well within in the system bulk, then these will agree with those using open boundary conditions. The value of the critical coupling $J_\mathrm{c}$ exhibits a weak correlation with the droplet size.
To elucidate this weak effect, we investigated three different settings: (a) free, (b) half-periodic, and (c) repulsive boundary conditions on various quantities. In Fig. \ref{fig:compare_rbc}, we plot, for square lattices of varying size, the difference between the original system and that arrived at by tuning the central bond $J_{i_0j_0} $ to the critical value $J_\mathrm{c}$. The energy difference between the initial and final ground states is computed with the same initial set of coupling constants (that with $J_{0}$). As seen in Fig. \ref{fig:compare_rbc}, for the largest system, there is no effect of the boundary conditions and all curves tend to a constant. 
However, if one tries to fit the data for the repulsive boundary conditions over all (including small) system sizes then it may seem that there is a non-trivial drop. For the open boundary condition
square, honeycomb, triangular, cubic, and bcc lattices,  
the average single bond excitation energies are, respectively, 1.9463, 1.7452, 2.3375, 2.2531, and 3.2386 (with standard deviations of 0.0047, 0.0042, 0.0056, 0.0096, and 0.0106).
As we explained earlier, when the value of $J_{i_0j_0}$ is varied, the associated change of the
ground-state energy (as evaluated with the original coupling) is either 
(A) $\Delta E=0$ or (B) $\Delta E= 2|J_0-J_\mathrm{c}|$. 

Thus, the only way in which a dependence on system size can arise is if $J_\mathrm{c}$ depends on the system size. In Table \ref{tab:parameters}, we provide, for different lattices of varying size, the Spearman correlation coefficients between the ZED boundary size $|\partial D|$ and $J_c$. 
As seen therein, the correlation between $J_\mathrm{c}$ and the droplet size is weak and does not naturally suggest a power-law nor other types of scaling. As Table \ref{tab:parameters} further makes clear however, for repulsive boundary conditions the latter correlation is nonetheless significantly larger than that present for half-periodic or open boundary conditions. Albeit very weak, the latter correlation illustrates that it is not unambiguous how to sample ``typical droplets'' for finite size systems. 

\begin{table}[bh]
	\centering
	\begin{tabular}{|c|c|c|c|c|c|}
		\hline
		& $L$ &$N_\text{sample}$& Correlation\\
		\hline
  honeycomb & 16&$10^5$&-0.0050\\
		\hline
		honeycomb& 96&$85395$ & 0.0027\\
		\hline
		square& 16&$10^5$&  -0.0010\\
		\hline
		square& 96&$10^5$ & -0.0019\\
		\hline
		square (HPBC) & 16&$10^5$ &0.0003\\
		\hline
		square (HPBC)& 96&$10^5$ & -0.003\\
		\hline
		square (RBC)& 8&$10^5$ & 0.0363\\
		\hline
		square (RBC) & 16&$10^5$& 0.0263\\
		\hline
		square (RBC)& 96&$10^5$ & 0.0075\\
		\hline
		triangular & 16&$10^5$ & -0.0064\\
		\hline
		triangular & 96&$10^5$ & -0.0019\\
		\hline
		simple cubic &10&$3987$ & -0.010\\
		\hline
		bcc & 7&$5950$ & -0.018\\
		\hline
	\end{tabular}
\caption{The Spearman correlation coefficients between
the ZED boundary area $|\partial D|$ and associated critical value of the coupling $J_\mathrm{c}$ at which a transition between ground states appears. We list the correlations for different types of lattices. Included are also two additional boundary conditions for the square lattice: the HPBC (Half-Periodic Boundary Condition) and the RBC (Repulsive boundary condition, see text). Note the correlation coefficient is significantly larger with RBC on small $d=2$ systems.}
\label{tab:parameters}
\end{table}

\section{Gauge symmetry and its implications} 
\label{evenJ}

In what follows, we illustrate two simple corollaries that stem from the well-known gauge symmetry that Ising spin-glass systems possess. In particular, we will demonstrate that any spin configuration (and thus any excitation of arbitrary energy) can be generated by applying a sequence of (generally non-critical and overlapping) ZEDs.

\subsection{The even nature of the probability distribution of coupling constants}
The probability distribution of the critical coupling strengths at which a ground-state transition occurs must be even. The proof of this assertion is immediate. With $i$ and $j$ denoting arbitrary nearest-neighbor sites and $t_{i} = \pm 1$, the gauge transformation
\begin{equation}
\label{gauge}
J_{ij} \to J'_{ij}=J_{ij}t_it_j,~~\sigma_{i} \to \sigma'_i=\sigma_i t_i
\end{equation}
leaves the general spin-glass Hamiltonian of 
Eq. (\ref{Hamiltonian:eq}) invariant.

We may set $t_{o}=-1$ (with the site ``$o$'' marking one of the two endpoints of the central bond) leaving $t_{r \neq o}=+1$ at all other sites $r$. One can partition the space $\mathcal{J}$ of the coupling constants over the entire lattice into two: $\mathcal{J}_>$ and $\mathcal{J}_<$, depending on whether the central bond is positive or negative. We may then establish a one-to-one mapping between these two spaces, as obviously the gauge transformation of Eq. (\ref{gauge}) is invertible. Additionally, an instance in $\mathcal{J}_<$, along with its `gauge mirror' in $\mathcal{J}_>$, must have the opposite signs of $J_\mathrm{c}$, according to the nature of the gauge transformation. Since $\mathcal{J}=\mathcal{J_<}+\mathcal{J_>}$, the distribution of critical couplings is even $P(J_\mathrm{c})=P(-J_\mathrm{c})$. In the main text, we employed the shorthand $P(|J_\mathrm{c}|)\equiv P(J_\mathrm{c})+P(-J_\mathrm{c})=2P(J_\mathrm{c})$.

\subsection{Excitations of arbitrary energy as composites of ZEDs} 
The gauge symmetry of Eq. (\ref{gauge}) implies that {\it any} Ising spin configuration  ${\cal{C'}}$ (which differs from the ground state ${\cal{C}}_{0}$ of the Hamiltonian $H$ of Eq. (\ref{Hamiltonian:eq})  by, say, ``$s$'' spin flips) can be cast as a ground state of spin-glass Hamiltonians $H'$  for some set of coupling constants. This fact is highly significant, as it allows us to construct any spin excitation of $H$ (of either finite or vanishing energy) as a geometric composite of ZEDs appearing in a sequence of Hamiltonians,
\begin{eqnarray}
\label{string-theory}
H \to H^{(1)} \to H^{(2)} \to \cdots \to H^{(f -1)} \to H'.
\end{eqnarray}
As we will explain, here, ``$f$'' denotes the total number of bonds that are flipped in sign in $H'$ relative to those in $H$. (For domain-wall excitations, this number of flipped bonds is the domain-wall surface area discussed in the main text.)  

To illustrate the claim embodied in \eqref{string-theory}, we may, e.g., start with the ground state ${\cal{C}}$ of Eq. (\ref{Hamiltonian:eq}). We then employ the gauge transformation of Eq. (\ref{gauge}) for a single site. Taking an arbitrary site ``$1$'' and setting its respective gauge variable to $t_1 =(-1)$ (with $t_{r} =1$ at all other lattice sites $r \neq \mbox{``}1\mbox{''}$) transforms the initial  Hamiltonian $H \to H^{(1)}$. By comparison to $H$, the couplings $J_{1p}$ (with $p$ denoting all nearest neighbor sites of ``$1$'') now change sign, $J_{1p} \to -J_{1p}$ in $H^{(1)}$. The ground state configuration of this new Hamiltonian differs from ${\cal{C}}$ by this single flipped spin (at site number ``$1$''). We may then iteratively proceed in such a manner to flip another Ising spin at an arbitrary site number ``$2$'' and then flip another spin, etc., so as to ultimately flip any set of spins in ${\cal{C}}$. Following any such gauge transformation at the $s'$-th step ($s' \le s$) in which in \eqref{gauge} only one single site field $t_{s'}$ is set equal to $(-1)$ with all others being one, the $z$ links (with $z$ denoting the lattice/graph coordination number) that are attached to site $s'$ flip their sign with all other links remaining untouched. The sign change of the former $z$ individual links may be further carried out sequentially in any order. Some of these links may be inverted multiple times as this gauge transformation is carried out site by site. Any even number of flips of a given link yields no change. In the final analysis, there is some number ($f$) of bonds that are flipped an odd number of times as the system evolves from ${\cal{C}}  \to {\cal{C}'}$; the full set of these flips realizes the gauge transformation between the two global spin configurations of ${\cal{C}} \to {\cal{C}'}$ (Eq. (\ref{string-theory})). Any individual flip of sign of a single link (say the $n-$th in \eqref{string-theory} (with $n \le (f-1)$)) generates a  transformation $H^{(n)} \to H^{(n+1)}$. This {\it single bond} transformation is precisely of the form that we investigated in the current work. As the Hamiltonians evolve according to \eqref{string-theory} (yielding the full multi-spin and link gauge transformation of \eqref{gauge}), the respective ground-states of change as 
\begin{eqnarray}
{\cal{C}} \to {\cal{C}}_{1} \to {\cal{C}}_{2} \cdots  \to {\cal{C}}_{f-1} \to {\cal{C'}}. 
\label{C-string-theory}
\end{eqnarray}

We next recall our previously established property (ii). This property implied that whenever ground-state spins change when an exchange-constant $J_{ij}$ on a given bond is varied to a new value $J'_{ij}$, these spins must belong to the ZED of the bond $(ij)$. Combining this with the fact that any spin configuration ${\cal{C'}}$ can be cast as the ground state of some spin-glass Hamiltonian $H'$, it follows as we will describe, a ``composition'' of the ZEDs appearing when individual exchange-constants $J_{ij}$ are sequentially changed will yield ${\cal{C'}}$. That is, after the completion of this process described by \eqref{string-theory} for all altered links $J_{ij}$, the set of spins that have been overturned a total odd number of times 
in the chain of Eq. (\ref{C-string-theory}) will form the spin excitation ${\cal{C}}'$ of the original system $H$. We now discuss any individual part ${\cal{C}}^{(n)} \to {\cal{C}}^{(n+1)}$ of the transformation chain of Eq. (\ref{C-string-theory})
that is associated with the change of the $n$-th link. Here, we note that similar to our earlier discussion of property (iii), the ground-state spin configuration ${\cal{C}}^{(n+1)}$ of $H^{(n+1)}$ is either \newline
(A) Equal to the ground-state ${\cal{C}}^{(n)}$ of $H^{(n)}$ 
(when the single exchange-coupling that is varied is not made to cross its critical value) or  \newline
(B) Forms a new ground state configuration which differs from ${\cal{C}}^{(n)}$ by the ZED of link $n$ (with the latter ZED delineated as it appears for the system defined by $H^{(n)}$). This arises  when a critical coupling crossing does occur as the single bond flips sign.
\newline
If overlaps exist between the sets of overturned spins (i.e., the ZEDs) between any of the $f$ individual steps of the transformation of Eq. (\ref{C-string-theory}) then no general statements may be made regarding the energies of general excitations ${\cal{C}'}$ (as evaluated with $H$) and their typical geometries for a given energy. However, it is worth noting that, statistically, {\it independent of any of the Hamiltonians of \eqref{string-theory}}, following each individual bond change, {\it the geometrical characteristics} of the flipped spins following each step adhere to the {\it universal} scaling relations and fractal dimensions for the volumes and surface areas that we reported on in this work. As noted in the main text, for the systems that we examined, our found fractal dimensions for single bond change ZEDs are very close to and statistically consistent with the fractal dimension of domain-walls that were generated by flipping a divergent number ($f = {\cal{O}}(L)$) of links at the system boundary.
In the thermodynamic limit, such asymptotically large domain-walls are composites of $f \to 
\infty$ (generally overlapping) individual ZEDs following the recipe  of \eqref{C-string-theory}. We must underscore that in determining the ZED fractal dimension, we investigated (inasmuch as possible numerically) the volume and surface area scaling of the asymptotically largest ZEDs (including any such ZEDs that might be present during individual steps in sequences such as that of \eqref{C-string-theory}, a sequence whose length diverges for asymptotically large domain-walls). 

Statistically, if the $f$ flipped bonds are far separated from one another with weak correlations amongst their respective individual ZEDs then each of the individual single bond excitation energies may be added. The latter follow the earlier discussed distribution $P(\Delta E')$  (see Fig. \ref{fig:jcdistribution} (a) with, for the square lattice, the average energy change $\langle \Delta E \rangle$ per bond of Fig. \ref{fig:compare_rbc}). For, e.g., such $f$ far separated bonds on a square lattice, the total mean excitation energy relative to the ground state of $H$ will be given by $f$ times the average square lattice  excitation energy $\langle \Delta E \rangle$ of Fig. \ref{fig:compare_rbc} for a single bond (i.e., by $\sim 2 f$ for the square lattice).  \bigskip

After the completion of this work, we were told by D. Stein that related ideas have been invoked in the study of excitations that are associated with a change of the boundary conditions \cite{newman_ground_2021} (in the discussion leading to Claim 9.5 therein). As emphasized above, the construct discussed in this Section applies to all configurations and thus to all excitations (boundary or otherwise).

\section{A comparison between our findings with the predictions of various theories}
\label{other_theories}

In what follows, we briefly discuss several key features of contending theories for excitations in the spin-glass phase and compare these with our results. For a detailed description of these theories, the reader is referred to some of Refs. \cite{parisi_infinite_1979,mcmillan_scaling_1984,FisherHuse88,BM,TNT1,palassini_nature_2000,Houdayer2000,newman_metastate_1997,newman_ground_2021}. Before presenting this short comparison, we would like to reiterate and underscore one of our central findings: \newline
\bigskip

{\it We find similar behaviors for ZEDs in both $d=2$ and $d=3$ lattices. That is, ZEDs exhibit similar features independent of whether ($d=3$) or not ($d=2$) a finite temperature spin-glass phase appears on lattices of dimension $d$.}\newline
\bigskip

$\bullet$ In the Replica Symmetry Breaking (RSB) picture, the ratio of the interface size to the volume tends to a non-vanishing constant in the thermodynamic limit \cite{parisi_infinite_1979}. In our investigations, we find (since the ZED surface fractal dimension is smaller than that of the system, $d_{\mathrm{s}} <d$) that the ratio of the number of spins lying on the ZED interface to the total number of spins in the system tends to zero with increasing system size. \newline
\bigskip

$\bullet$ In the droplet picture  \cite{mcmillan_scaling_1984,FisherHuse88,BM}, the lowest energy excitations of linear size $\ell$ are compact. The associated excitations have a volume
(i.e., that associated with the total number of flipped spins) fractal dimension equal to the system spatial dimension $d$ and a surface fractal dimension $d_f$ satisfying $d-1< d_{\mathrm{f}}<d$. The latter inequality would be consistent with our findings for ZEDs if one sets the latter predicted surface fractal dimension to that of the ZEDs that we investigated, i.e., following the substitution 
$d_\mathrm{f} = d_\mathrm{s}$. Moreover, we cannot certify, within our finite size numerical analysis, that the volume fractal dimension is equal to the spatial dimension $d$ of the lattice.

Additionally, according to the droplet theory \cite{mcmillan_scaling_1984,FisherHuse88,BM}, for droplets of linear scale $\ell$, the droplet energies scale as $\Delta E \propto \ell ^\theta$, with a constant density of states near vanishing energy excitations. 
We now turn to our results. Since ZEDs are, by definition, excitations of vanishing energy {\it for any $\ell$}, one cannot simply define a meaningful finite stiffness exponent $\theta$ to describe the relation between ZED energies and their geometrical linear size. It is possible that the ZEDs do not constitute the ``typical'' lowest energy excitations of a given $\ell$ in the droplet theory. Furthermore, as we described in detail, the ground state of a bond configuration associated with the change of a single bond constitutes an excitation of the original bond configuration (i.e., that before the flip of the single bond). As we illustrated, our numerical results show that, for typical boundary conditions, the latter energy cost is independent of the linear system size $L$ (and the average droplet size $\ell$ which scales with $L$) when $L$ becomes large.  
\newline
\bigskip

$\bullet$ The trivial-non-trivial (TNT) theory \cite{TNT1,palassini_nature_2000,Houdayer2000} posits that either one of the droplet or the RSB theories may capture various aspects of the physical phenomena. The TNT theory suggests that in $d=3$ dimensions a finite fraction of the flipped spins 
in the critical droplet will be on the droplet boundary while in $d=2$ lattices that fraction tends to zero.\newline
\bigskip

$\bullet$ Lastly, 
in the chaotic-pair (CP) picture \cite{newman_metastate_1997}, the excitations are, ``space-filling'' \cite{newman_ground_2021} (i.e., the ratio of the number of bonds lying on the excitation interfaces relative to the total number of bonds tends to a finite number in the thermodynamic $L\to \infty$ limit). That is, in the CP scenario, the surface fractal dimension of the excitation is equal to that of the system, $d_\mathrm{f} =d$. In our investigation, we find that for the ZEDs, the interface surface fractal dimension $d_{\mathrm{s}} <d$. If we set, for comparison, $d_{\mathrm{f}} = d_{\mathrm{s}}$ then our results will imply that $d_{\mathrm{f}}<d$.

\end{document}